\RequirePackage[hyphens]{url}
\documentclass[manuscript,screen,nonacm]{acmart}
\usepackage{tikz}

\usepackage{amsmath,amsfonts}
\usepackage{graphicx}
\usepackage{textcomp}
\usepackage{xcolor}
\usepackage{lipsum}

\usepackage{multirow}
\usepackage[ruled,vlined]{algorithm2e}
\usepackage{algpseudocode}

\usepackage{array}
\usepackage{booktabs}
\usepackage{threeparttable}
\usepackage{color, colortbl}
\usepackage{xcolor}
\usepackage{tcolorbox}
\usepackage{xparse}
\usepackage{tabularx}
\usepackage{siunitx}

\definecolor{answerboxbg}{RGB}{248, 249, 250}
\definecolor{answerboxborder}{RGB}{206, 212, 218}
\definecolor{answerboxlabel}{RGB}{13, 110, 253}

\NewDocumentEnvironment{answerbox}{O{} m}{%
    \begin{tcolorbox}[
        colback=answerboxbg,
        colframe=answerboxborder,
        boxrule=0.8pt,
        arc=3pt,
        left=8pt,
        right=8pt,
        top=8pt,
        bottom=8pt,
        before skip=10pt,
        after skip=10pt,
        fonttitle=\bfseries\sffamily\small,
        coltitle=answerboxlabel,
        title={\ifx\relax#1\relax Answer\else #1\fi}
    ]
    \textbf{#2}
    
    \vspace{4pt}
    \end{tcolorbox}
}

\newcommand{\takeaway}[2]{%
  \begin{tcolorbox}[
    colback=lightgray!20, 
    boxrule=0.2pt, 
    arc=1pt,
    left=0mm, 
    right=0mm, 
    top=0mm, 
    bottom=0mm,
  ]
  \textbf{RQ#1}: #2
  \end{tcolorbox}
}

\usepackage{bm}

\usepackage[tight,footnotesize]{subfigure}

\usepackage{url}

\usepackage[svgnames]{xcolor}
  \definecolor{diffstart}{named}{Grey}
  \definecolor{diffincl}{named}{Green}
  \definecolor{diffrem}{named}{OrangeRed}

\usepackage{listings}
\usepackage{courier}

\definecolor{mygreen}{rgb}{0,0.6,0}
\definecolor{mygray}{rgb}{0.5,0.5,0.5}
\definecolor{mymauve}{rgb}{0.58,0,0.82}

\lstdefinelanguage{JavaScript}{
  morekeywords=[1]{break, continue, delete, else, for, function, if, in,
    new, return, this, typeof, var, void, while, with},
  morekeywords=[2]{false, null, true, boolean, number, undefined,
    Array, Boolean, Date, Math, Number, String, Object},
  morekeywords=[3]{eval, parseInt, parseFloat, escape, unescape},
  sensitive,
  morecomment=[s]{/*}{*/},
  morecomment=[l]//,
  morecomment=[s]{/**}{*/}, 
  morestring=[b]',
  morestring=[b]"
}[keywords, comments, strings]

\usepackage{enumitem}
\setlist[itemize]{noitemsep, topsep=0pt}
\setlist[enumerate]{noitemsep, topsep=0pt}

\newif\ifdraft
\draftfalse 

\ifdraft
    \newcommand\ja[1]{{\color{orange}[JA]: #1}}
    \newcommand\og[1]{{\color{green!60!black}[OG]: #1}}
    \newcommand\kr[1]{{\color{purple}[KR]: #1}}
\else
    \newcommand\ja[1]{}
    \newcommand\og[1]{}
    \newcommand\kr[1]{}
\fi

\newif\ifarxiv
\arxivtrue      

\newcommand{\arxivonly}[1]{%
  \ifarxiv
    #1%
  \fi
}

\newcommand{\reviewonly}[1]{%
  \ifarxiv
  \else
    #1%
  \fi
}

\arxivonly{
  \setcopyright{none}
  \settopmatter{printacmref=false}
  \renewcommand\footnotetextcopyrightpermission[1]{}
}


\begin{document}
%
\title{GitHub Engagement Signals for CVE Prioritization: The GitHub Popularity Metric (GPM)}


\author{Jafar Akhoundali}
\affiliation{%
  \department{LIACS}
  \institution{Leiden University}
  \city{Leiden}
  \country{Netherlands}
}
\email{j.akhoundali@liacs.leidenuniv.nl}

\author{Kristian Rietveld}
\affiliation{%
  \department{LIACS}
  \institution{Leiden University}
  \city{Leiden}
  \country{Netherlands}
}
\email{k.f.d.rietveld@liacs.leidenuniv.nl}

\author{Olga Gadyatskaya}
\affiliation{%
  \department{LIACS}
  \institution{Leiden University}
  \city{Leiden}
  \country{Netherlands}
}
\email{o.gadyatskaya@liacs.leidenuniv.nl}

\begin{abstract}
Attackers can compromise multiple systems with a single vulnerability, while defenders need to fix all security weaknesses in their systems. This asymmetry puts defenders at a disadvantage. Security vulnerabilities are found at an alarming rate, and patching vulnerabilities is costly and time-consuming; thus, vulnerability prioritization is a must and a time-critical challenge. Many prioritization metrics, such as the CVSS, EPSS, KEV, and SSVC, are currently used, each with different pros and cons, such as openness, degree of automation, time-criticality, coverage, and need for expert input.

In this work, we propose the GitHub Popularity Metric (GPM), a fully open, publicly computable prioritization metric based on the popularity of exploits in GitHub repositories. We use GitHub features such as the number of stars, forks, and related unique users to create a metric that indicates the popularity of CVEs across different time frames, both relative to the current time and historically.
We compare the proposed metric with various existing vulnerability prioritization metrics and known exploited vulnerabilities and demonstrate that it provides tangible insights for defenders, identifying unique CVEs and inconsistencies in existing methods. The GPM was integrated into the EPSS version 5. 
\reviewonly{We share our code, allowing full reproducibility. We also implemented an online platform to showcase the GPM and inform developers and defenders.}

\noindent\textbf{Note.}
A demonstration based on this work has been accepted to the demo track of the
ACM Conference on Computer and Communications Security (CCS) 2026~\cite{gpm2026}.
\end{abstract}

\maketitle

\pagestyle{plain}

\section{Introduction}\label{sec:introduction}
The volume of software development continues to increase each year, as reflected in the number of software vulnerabilities. Common Vulnerabilities and Exposures (CVE) is a standard for making reference identifications of vulnerabilities. As of the time of writing, on average, 212 CVEs are published per day, with an annual growth rate of 21.1\%~\cite{cveicu}.
The increase in CVE volume gives attackers the upper hand and makes defenders' task of compromising a system ever more challenging; sometimes a single vulnerability is sufficient, but defenders need to patch all vulnerabilities. 
Moreover, not all vulnerabilities are equal for attackers, as their exploitability, ease of exploitation, impact, and prevalence differ. In fact, only a fraction of published CVEs will ever be exploited~\cite{mell2024measuring,sabottke2015vulnerability,shimizu2025vulnerability}. On the defenders' side, fixing vulnerabilities is costly. The large number of CVEs, the cost of identifying and fixing them within one's own infrastructure, and the urgency to patch vulnerabilities before they are exploited require the triage of CVEs to be patched. The discussed asymmetry between attackers and defenders necessitates patch prioritization methods that enable defenders to respond to risks more quickly and intelligently.  

The Common Weakness Enumeration (CWE) tags and the Common Vulnerability Scoring System (CVSS) metric are often included with CVE descriptions. The CVSS was defined as a metric to assess the severity of vulnerabilities. However, it has several deficiencies, including the need for substantial manual expert effort to label its parameters and the problematic prioritization of CVE patching based on the CVSS. For example, the CVSS does not correlate with exploited vulnerabilities in the wild~\cite{howland2023cvss}, and not only can multiple people disagree on a CVSS score, but also the same person may give different scores months apart~\cite{wunder2024shedding}. 

Apart from the CVSS, several reputable metrics and threat intelligence sources for vulnerability prioritization have been proposed in the academic literature and industry~\cite{jacobs2021exploit,jiang2025survey,shrestha2020multiple,patel2024harnessing,elder2024survey,yang2020better,mell2025likely}. However, each has its own deficiencies. For example, the EPSS~\cite{jacobs2023enhancing,jacobs2021exploit} tends to underestimate the chance of vulnerability exploitation~\cite {mell2025likely}, and some of its data sources are not transparent. Reputable exploited-vulnerability lists, such as CISA KEV or Vulncheck KEV, may experience substantial delays or limited visibility into the addition of actually exploited vulnerabilities due to the manual effort required. Another important research direction has been to use social media signals from platforms like X (Twitter) and Reddit to predict vulnerability exploitation~\cite{shrestha2020multiple,chen2019using,zong2019analyzing,alves2020follow,horawalavithana2019mentions,sabottke2015vulnerability,schiappa2019cyber,de2020evaluating}. However, the utility that these social media platforms provide in terms of the vulnerability exploitation signals diminished in the past few years due to the implemented restrictions in the public API access. 


In this work, we introduce a novel \textbf{GitHub Popularity Metric} (\textbf{GPM}) for prioritizing CVEs, which is based on interactions with public GitHub repositories hosting exploit proof-of-concept repositories. It provides a transparent, non-binary (numeric) value based on historical data that helps prioritize CVEs by popularity, which can be especially advantageous for defenders. Moreover, the GPM relies only on publicly available data from GitHub and can be computed by anyone for free. 

Significantly, GPM does not rely on post-exploitation activity-detection sources and therefore does not incur the associated delays. We show that insights from the GitHub Popularity Metric provide predictive value \emph{both as an independent signal} and \emph{as a complement to existing metrics}, improving vulnerability prioritization for patch management. Following the EPSS~\cite{jacobs2023enhancing}, we compare CVE exploitation status with the corresponding GPM scores. Specifically, we address the following research questions:
\begin{itemize}[leftmargin=9pt]
    \item \textbf{RQ1:}  How popular is GitHub for sharing exploits in comparison with other platforms?
    \item \textbf{RQ2:}  What is the delay between the appearance of public GitHub exploits and the publication of their corresponding CVEs?
    \item  \textbf{RQ3:}  What is the relationship between public CVE exploits in GitHub and known exploited CVEs?
    \item \textbf{RQ4:}  How effective is prioritizing CVEs based on public exploits on GitHub, in comparison with traditionally used sources?
    \begin{itemize}
        \item \textbf{RQ4.1:} How much does the GPM align and complement the established prioritization metrics?
        \item \textbf{RQ4.2:} Is the GPM able to identify relevant yet overlooked CVEs?
    \end{itemize}
    \item \textbf{RQ5:}  What are the advantages and disadvantages of using the GPM over existing approaches?
\end{itemize}

\textbf{Contributions.} In this paper, we introduce the GitHub Popularity Metric and show its value in comparison to other established vulnerability data sources and prioritization approaches. We demonstrate the actionable insights GPM provides to defenders by uncovering critical exploited CVEs not covered in other sources and highlighting inconsistencies in established approaches. 

We demonstrate that the GPM is predictive in uncovering exploited CVEs rather than reactive to the addition of a CVE to CISA's list of key exploited vulnerabilities. The GPM proposes tangible improvements to SSVC (Stakeholder-Specific Vulnerability Categorization). It aligns strongly with the industry standard for exploitation prediction, the EPSS, while also highlighting unique CVEs not visible in other reputable sources. Our work has been merged into the EPSS version 5 (in a modified way)\footnote{\url{https://www.first.org/epss/how-it-works}} and has resulted in MITRE assigning published status to several CVEs uncovered by the GPM. We share our code and data and provide a web app that may inform defenders of the GPM values.
\reviewonly{
\footnote{\url{http://45.14.224.176/} Please refer to Open Science section for credentials}
}
\section{Background and Related Work}\label{sec:background}
In this section, we introduce the key concepts about vulnerabilities and the existing reputable systems that help prioritize vulnerabilities. When referring to the amount of data available at a particular system or database, we provide the numbers collected at the time of developing the GPM (February 2026), unless explicitly stated otherwise. 

\subsection{Vulnerabilities and Exploits}\label{sec:cves}
\noindent \textbf{Vulnerabilities and CVEs.}
A \emph{vulnerability} is a weakness that allows unintended actors to affect the integrity, confidentiality, or availability of a system unexpectedly. The Common Vulnerabilities and Exposures (CVE) system is a standard, publicly disclosed catalog of known vulnerabilities that enables the identification and tracking of vulnerabilities discovered in commonly used systems across various entities and tools. The weakness category of the CVEs is labeled with the Common Weakness Enumeration (CWE) system.  MITRE hosts the CVE Secretariat and maintains the program infrastructure.\footnote{\url{https://www.cve.org/}} It also gives different parties, such as vendors, projects, and national CERTs, the authority to publish CVEs within their defined scope. Such entities are called CVE Numbering Authorities (CNAs)~\cite{cvemitre}; there were 506 of them as of February 2026.
 
\noindent \emph{CVE lifecycle.} When a vulnerability is discovered, it is first reported to a CNA, and the CNA assigns a CVE ID to it. A CVE ID (commonly referred to simply as a CVE) initially has the ``RESERVED'' status. After confirming the issue and filling in the required information, such as affected products, fixed version, a public reference, etc., the CVE is made public and gets the ``PUBLISHED'' status. CVEs may have other statuses. CVEs with the ``REJECTED'' status mean they were assigned incorrectly, the CNA took them down, or they were assigned as duplicates; thus, a general recommendation is to avoid using rejected CVE IDs. 
 
\noindent \textbf{NVD.}
The National Vulnerability Database\footnote{\url{https://nvd.nist.gov/}} (NVD) is an important source of CVE data managed by the US National Institute of Standards and Technology (NIST). In addition to the standard CVE lifecycle statuses, the NVD provides extra tags to ``PUBLISHED'' CVEs. For example, published CVEs in the NVD can have a status of ``RECEIVED'', ``AWAITING ANALYSIS'', ``UNDERGOING ANALYSIS'', ``ANALYZED'', ``MODIFIED'' and ``DEFERRED''. For details about each status in the NVD, we refer to the NIST website~\cite{nist_nvd_vulnerability_status_2025}. 

Moreover, in the NVD, CVEs are enriched with additional information that helps defenders to prioritize CVEs for patching and understand their impact better. This additional information includes, for example, a CWE tag, a list of the affected software or hardware components (the Common Platform Enumeration (CPE) strings), and other details. However, the high volume of new CVEs and the recent lack of funding have introduced a huge delay in reviewing CVEs and adding valuable information to the NVD. In 2025, NIST announced that it will not enrich vulnerabilities issued before 2018, assigning them the status of ``DEFERRED''~\cite{nist_nvddefered}. In 2026, they announced that due to the high volume of CVEs, they will only enrich CVEs that are included in KEV, software used in the federal government, and a list of critical software~\cite{nist_nvd_update_2026}. At the time of our data collection, out of the total 332,207 CVEs, the NVD marked 94,568 CVEs as ``DEFERRED'', meaning they are not prioritized for enrichment; however, this does not mean such CVEs are no longer dangerous or exploited. Moreover, 975 CVEs were under review, and 28,394 CVEs were awaiting analysis~\cite{nvd-dashboard}. 
 
\noindent \textbf{Exploit sources.} There are diverse public and private data sources of CVE exploits. Unfortunately, accessing private data sources, such as exploitation evidence, is challenging; thus, we consider only publicly available data. Exploit-DB\footnote{\url{https://www.exploit-db.com/}} is a community-driven service that publishes exploits received from the community. Not all exploits published there are associated with CVEs, and some reviewers can verify exploits. Metasploit\footnote{\url{https://www.metasploit.com/}} is an open-source framework used by security professionals that contains several modules, including exploit modules, payloads, and encoders. Although fewer exploits are available in Metasploit (\num{2628}) than in Exploit-DB (\num{46991}), Metasploit provides a more uniform, well-designed exploitation framework. It is worth noting that many of these exploits may be duplicates for the same CVEs, and not all of them are assigned CVE tags.
Nuclei\footnote{\url{https://projectdiscovery.io/}} is a well-known open-source HTTP-based vulnerability scanner that provides a customizable framework to create security scanning templates. Thanks to the convenient YAML-based template formatting and active community, it currently contains \num{12644} templates. 

Although GitHub is not explicitly designed to share exploits, it has gradually become a popular source to discuss and share exploits~\cite{elyadmani2025securepoc}. GitHub has previously removed certain exploits from its platform, such as ProxyLogon. However, based on feedback from the security community~\cite{github_aupolicy_pr397_2021}, they decided to change this and allow dual-use security content on GitHub, but may remove the material when GitHub is used as an exploit or malware content delivery network (CDN)~\cite{githubexploit}. The main benefit of this approach is that, while exploits are already traded privately among threat actors, restricting their public availability can disadvantage defenders by limiting their ability to develop patches. So far, this policy has been broadly accepted, and exploit repositories and tools continue to be shared via GitHub~\cite{elyadmani2025securepoc}. We note that both Metasploit~\cite{rapid72025metasploit} and Nuclei~\cite{projectdiscovery2025nuclei}, and many active GitHub repositories~\cite{trickest2025cve,qazbnm4562025awesome}, collect numerous exploits. 

\subsection{Reputable CVE Prioritization Sources}\label{sec:datasources}
This section summarizes the key data sources on vulnerability severity and exploitability that practitioners use to prioritize CVEs for patching.
 
\noindent \textbf{CVSS.}
The Common Vulnerability Scoring System (CVSS), developed by the Forum of Incident Response and Security Teams (FIRST), is an open standard to assign a numeric severity score to a vulnerability. 
In practice, CVSS does not incorporate rich exploitation data, which limits its effectiveness for vulnerability prioritization. Although the latest version, CVSS v4.0, introduces an exploit maturity metric, it has not yet seen widespread adoption. Additionally, studies have shown that the CVSS is not consistent~\cite{wunder2024shedding} and is not very effective for vulnerability prioritization because it does not reflect the actual risk of a vulnerability's exploitation in practice~\cite {cvss_fallacy_blackhat_2022, howland2023cvss, cvss_dos_blackhat_2013, zhang2023flaw, scarfone2009analysis, spring2021time, buying_into_bias_blackhat_2013,massacci2024holy,tan2025analysis}. Suciu et al. utilized an exploitability sub-score of CVSS~\cite{nist-cvss-v31-equations} and demonstrated that the exploitability of CVSS versions\footnote{At the time of writing this paper, CVSS version 3 is the most used version.} 2 and 3 had a maximum precision of ca. 0.19, meaning over 80\% of vulnerabilities flagged as ``likely
to be exploited'' are actually never exploited~\cite{suciu2022expected}.
 
\noindent \textbf{EPSS.} 
FIRST developed the Exploit Prediction Scoring System\footnote{\url{https://www.first.org/epss/}} (EPSS), a machine learning-based approach to predict the probability of exploitation of a CVE in the next 30 days~\cite{jacobs2021exploit}. The EPSS is calculated automatically daily and is currently well-recognized and respected in the industry, as it is publicly available and uses hundreds of variables for calculation. As the EPSS provides a numeric value between 0 and 1 (probability), it allows CVE prioritization. However, it also has notable disadvantages. The EPSS uses an open methodology, but the data used in the pipeline are not fully public, and the model itself is black-box and non-explainable~\cite{jacobs2023enhancing}. Although the EPSS factors in the availability of exploits on GitHub, it does not consider the exploit's popularity, and it is not known what exactly is included in the EPSS score about this factor. Historically, EPSS included X (Twitter) mention counts as a numeric signal, but with the loss of X’s free API in 2023, it seems this metric is no longer used. Currently, the EPSS considers CVE exploitation evidence in a somewhat restrained way: its data is limited by manual analysis reports and the availability of data from honeypots, as well as the development of detection rules for the considered CVE\footnote{This concerns EPSS version 4, as in EPSS‌ version 5, the results of this work are integrated as a data source}.  
By nature, the EPSS data sources might be biased toward detecting network-based attacks instead of, for example, attacks requiring physical proximity, as they are harder to detect. Furthermore, the EPSS only works on fully accepted and published CVEs. All these limitations increase the delay in detecting actually exploited CVEs and bias the EPSS score towards certain CVEs commonly detected in exploitation activities.
 
\noindent \textbf{CISA KEV.} 
The Known Exploited vulnerabilities\footnote{\url{https://www.cisa.gov/known-exploited-vulnerabilities-catalog}} (KEV) list was initiated on 2021-11-03 by the US CISA. This is a regularly published list of vulnerabilities in software, hardware, and firmware, so long as they are known to be actively exploited. Currently, the US federal civilian agencies are required to remediate the KEV-listed vulnerabilities by a specified due date, while globally, the CISA KEV list is publicly used as a best practice to patch vulnerabilities. Inclusion in KEV usually requires credible evidence of exploitation and is typically based on reliable government, industry, or open-source threat intelligence. Despite these advantages, the CISA KEV has certain limitations for vulnerability prioritization. First, its coverage is low, only \num{1513} CVEs. 
For example, Koscinski et al.~\cite{koscinski2025conflicting} recently discussed the lack of consistency between CVE metrics such as the EPSS and CVSS, which can result in misalignment and confusion. Their analysis was limited to 600 CVEs from Microsoft Patch Tuesday releases, which, when compared with CISA KEV, resulted in an intersection of only 13 exploited CVEs.
Second, updates are processed manually and are typically published only on US working days; vulnerabilities discovered/exploited on weekends can have a two- to three-day publication delay~\cite{garrity_vulncheck_2024}. In our dataset, there is only one CVE added on Sunday (CVE-2025-53770~\cite{cisa_toolshell_alert_2025}). Finally, as a list, KEV is binary and does not provide a ranking of CVEs. 
 
\noindent \textbf{KEV-V.}
A private threat intelligence provider, VulnCheck, actively maintains a list of independently identified, actively exploited vulnerabilities, similar to CISA KEV. VulnCheck KEV\footnote{\url{https://www.vulncheck.com/kev}} (KEV-V) is a superset of exploited CVEs detected in various sources, including Shadowserver\footnote{\url{https://www.shadowserver.org/}} and CISA KEV; thus, it offers a large coverage (\num{4599} CVEs). At the same time, as this is a private data source, it is not fully transparent, and its users depend on VulnCheck. 
 
\noindent \textbf{SSVC.} 
CVE descriptions often suffer from data incompleteness, especially in the exploitation details~\cite{guo2020predicting,gonzalez2019automated,croft2023data}. In 2024, CISA introduced the ADP process\footnote{\url{https://www.cve.org/Media/News/item/blog/2024/06/04/CISA-Added-as-CVE-Authorized-Data-Publisher}}, which enriches previously issued CVE records with fields like vulnerability type, description, impact, and exploitation. An interesting component that CISA introduced through the ADP is called Stakeholder-Specific Vulnerability Categorization (SSVC)\footnote{\url{https://www.cisa.gov/stakeholder-specific-vulnerability-categorization-ssvc}}, which provides data related to exploit availability and evidence of exploitation~\cite{spring2021prioritizing}. SSVC is another reputable data source for prioritizing CVEs. Summaries of the SSVC fields are shown in Table~\ref{tab:ssvc_values}. 

\renewcommand{\multirowsetup}{\centering}
\begin{table}[t]
\centering
\renewcommand{\arraystretch}{1.2}
\setlength{\tabcolsep}{3pt}
\caption{CISA SSVC Possible Values and Descriptions}
\label{tab:ssvc_values}
\begin{tabular}{
    |>{\centering\arraybackslash}m{1.9cm}|
    >{\centering\arraybackslash}m{1.0cm}|
    >{\centering\arraybackslash}m{4.5cm}|
}
\hline
\textbf{Field} & \textbf{Value} & \textbf{Description} \\
\hline
\multirow{3}{2.1cm}[-2ex]{\centering Exploitation} 
    & None   & No evidence of exploitation; no PoC available \\
\cline{2-3}
    & PoC    & Public PoC code or exploitation method available \\
\cline{2-3}
    & Active & Reliable evidence of in-the-wild exploitation \\
\hline
\multirow{2}{2.1cm}[-1.2ex]{\centering Automatable} 
    & No     & Requires manual effort or major customization \\
\cline{2-3}
    & Yes    & Can be reliably automated (e.g., RCE, command injection) \\
\hline
\multirow{2}{2.1cm}[-1ex]{\centering \shortstack{Technical\\Impact}} 
    & Partial & Partial control or limited data disclosure (e.g., DoS) \\
\cline{2-3}
    & Total   & Full system control or total data disclosure \\
\hline
\end{tabular}
\end{table}
 
\noindent \textbf{AlienVault Open Threat Exchange (OTX) CVEs.} 
AlienVault OTX\footnote{\url{https://otx.alienvault.com/}} is a free, community-driven threat intelligence platform.
OTX organizes information in pulses, which are collections of related threat intelligence (e.g., indicators of compromise, malware samples, or attack techniques). Community members can map pulses to vulnerabilities and CVEs, thereby linking threat intelligence to the specific security issues. Pulses can be a valuable indicator that a vulnerability is exploited. While checking all thousands of CVE-related pulses is not practical, we manually analyzed CVEs with an exploit on GitHub: we collected available pulses for these CVEs and manually checked them, resulting in the discovery of 14 exploited CVEs that were not in KEVs as of Aug 2025. 
 
\noindent \textbf{The ransomware dataset.}
Ransomware groups frequently exploit CVEs in their operations. Using publicly available threat intelligence data~\cite{BushidoUK2025VulnerabilityMatrix}, we collected a list of CVEs known to be exploited by ransomware threat actors. 
 
\noindent \textbf{Unavailable sources.}
We note that these latter two data sources (the OTX CVEs and the ransomware dataset) represent a common approach in industry, when CVE prioritization is based on reliable exploitation signals from reputable threat intelligence platforms. In this space, the Shadowserver Foundation is one of the major players in providing threat intelligence insights to the public, including scanning and notifying vulnerable systems~\cite{shadowserver2025geovision}. GreyNoise, another well-known source of threat intelligence that provides actionable insights, recently discovered a zero-day vulnerability in live-stream cameras using LLMs~\cite{stone2024greynoise}. Moreover, GreyNoise mentioned that spikes in malicious activities usually precede the disclosure of new CVEs~\cite{greynoise2025earlywarning}. However, we do not have access to either Shadowserver or GreyNoise data. Further, we attempted to get the artifacts of the expected exploitability study~\cite{suciu2022expected}, but did not get a response. 

\subsection{Related Work}\label{sec:relwork}
\noindent \textbf{Vulnerability patch prioritization.} 
Due to the exponential growth of CVEs, prioritization becomes a central operational challenge for defenders~\cite{RomanoskyJacobs2022EPSSPercentiles}.
 
\noindent \emph{Social media signals.}
Several studies used CVE mentions in online platforms like X (formerly Twitter), Reddit, or GitHub as early signals and sources to identify which vulnerabilities will be exploited in the wild~\cite{shrestha2020multiple,chen2019using,zong2019analyzing,alves2020follow,horawalavithana2019mentions,sabottke2015vulnerability,schiappa2019cyber,de2020evaluating}. Such models benefit defenders and help them focus on truly dangerous CVEs early. Subsequently, free online services such as CVETrends\footnote{\url{https://cvetrends.com/}} were created to flag CVEs receiving a lot of social media attention. Unfortunately, in February 2023, X deleted its free API tier, resulting in less visibility on social media~\cite{xremovedfreetier}. In 2020, Shrestha et al. compared how vulnerabilities are discussed across Twitter, GitHub, and Reddit~\cite{shrestha2020multiple}. They mentioned that vulnerabilities mostly start on GitHub (46\% of discussions). Shrestha et al. mentioned ``\textit{We found GitHub is the most useful platform, but due to the lack of a public API, actionable insights can still be gathered from Twitter and Reddit}''. Similar conclusions were reached in~\cite{schiappa2019cyber,neil2018mining}. In July 2023, Reddit also limited public API usage. Fortunately, currently, GitHub allows searching for repositories. With a free, authenticated user, actionable insights can be collected in a reasonable time with some optimizations; however, rate limits still apply. This is further discussed in Sec.~\ref{sec:data}. 
 
\noindent \emph{Vulnerability prioritization frameworks.}
Vulnerability prioritization is an active research topic. A recent survey by Jiang et al.~\cite{jiang2025survey} outlines the key research directions, summarizes the different metrics like severity and exploitability used for prioritization, and summarizes the key literature studies on this topic. Elder et al.~\cite{elder2024survey} recently surveyed the vulnerability exploitability assessment literature. They observed that the CVSS, while lacking contextualization, still dominates vulnerability prioritization. However, hybrid models that combine the CVSS with dynamic features show promising results. Moreover, exploitability metrics evolved from static metrics to dynamic metrics such as exploit availability. Yet, potentially poisoned data and misinformation in social media, and a lack of proper integration of various sources remain open problems. 
Threat intelligence feeds have been explored for predicting vulnerability exploitation~\cite{greynoise2025earlywarning,patel2024harnessing}. 
Seker and Meng proposed combining signals from social media, GitHub, the dark web, and CVSS scores, and prioritizing CVE patching based on the final calculated scores for risk-based prioritization~\cite{seker2023xvrs}. Yang et al.~\cite{yang2020better} predicted the exploitability of CVEs based on the existence and source of a PoC. NIST has recently proposed a metric to predict likely exploited vulnerabilities (LEV), which is meant to augment the EPSS~\cite{mell2025likely}. In contrast, the GPM derives signal from the engagement dynamics of GitHub repositories, capturing community attention as a continuous, time-varying metric rather than treating PoC availability as a binary feature or combining signals from private sources.

\noindent \textbf{Vulnerability exploits.}
Not all CVEs are exploited, and not all CVEs are verified vulnerabilities. A recent survey study of researchers reported that 54\% of participants incorrectly believe that a CVE is assigned only if the vulnerability is verified~\cite{schloegel2025confusing}. This significantly compromises judgment for prioritizing patches. Moreover, even verified CVE vulnerabilities are not necessarily exploited. Khoury et al.~\cite{khoury2025mining} analyzed recent CVE exploits and reported that, of 111,331 CVEs published over 5 years, only 4,008 CVEs were exploited (ca. 3.6\%). The report is also consistent with other studies of the exploited ratio between 2\% and 5\%, following a stable ratio over the years~\cite{sabottke2015vulnerability,brilhante2023measuring}.
Kita et al. investigated the prevalence of exploit code repositories in GitHub, and compared it with ExploitDB and Metasploit~\cite{kitaprioritization}. They found that GitHub exceeds other sources in terms of coverage, real-world exploits, and timeliness. They mostly focus on the credibility and trustworthiness of exploits, while the GPM (our proposed method) is concentrated on popularity. Suciu et al.~\cite{suciu2022expected}, Householder et al.~\cite{householder2020historical}, and Iannone et al.~\cite{iannone2024early} investigated the historical perspectives of CVE exploitability, how CVE exploits become functional, and proposed methods for predicting exploitability in the future. In contrast, Al Asadi et al.\cite{al2022no} studied the lifecycles of the Internet-of-Things exploits.

\noindent \textbf{GitHub mining.}
Malicious actors also abuse GitHub and publish fake exploits, targeting developers and security engineers. When used, the exploit compromises the victim's system in return~\cite{elyadmani2025securepoc}. Such exploits can be removed if reported or seen by GitHub staff~\cite{he20244}.
GitHub stars have also been explored as a proxy to identify high-quality repositories~\cite{munaiah2017curating}. A star classifier was discovered to have extremely high precision (96\%) but low recall (27\%). Other studies demonstrate that there are active malicious players that use stars, forks, or pollute GitHub to spread malware~\cite{tania2024creating,he20244,Cao2022WhatTF,rokon2020sourcefinder,gonzalez2021anomalicious,qian2022malicious}. However, since GitHub works to remove fake stars, accounts, and malicious projects, their long-term impact is limited~\cite{he20244}. 
\section{Dataset}\label{sec:data}
In this section, we outline the data sources we used in this work, describe our data collection process, provide some statistics on the data, and answer RQ1 and RQ2.

\begin{figure*}[t]
  \centering
  \includegraphics[width=1.0\textwidth, keepaspectratio]{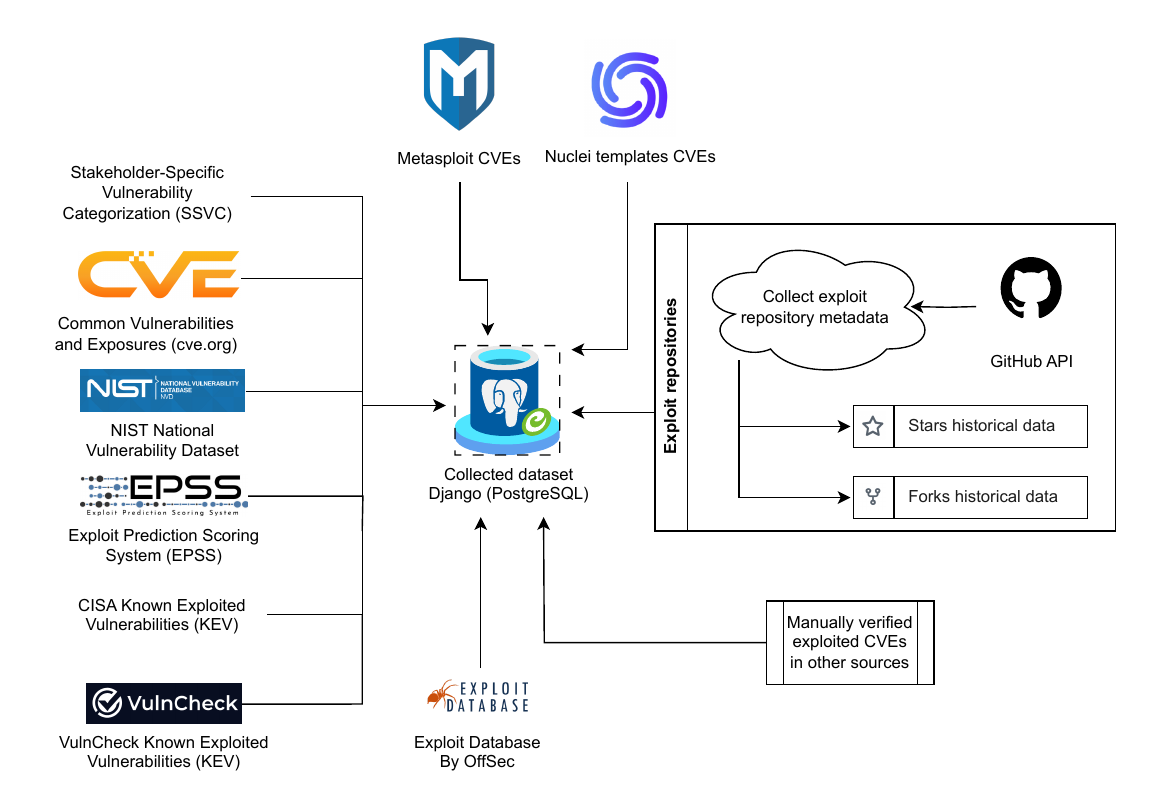}
  \caption{An overview of data sources that are collected into our dataset.}
   \label{fig:workflow}
\end{figure*}

\subsection{Data Sources}
To collect CVEs, we used the MITRE \url{cve.org} data, which also includes the SSVC. We scraped exploits from GitHub using the GitHub Search API and retrieved the GitHub star and fork history, including user IDs. We also collected daily EPSS data. 
Unlike similar works, we do not focus solely on accepted CVEs. Our motivation is that, in many cases, CVEs may take time to be published, and occasionally they may even be rejected incorrectly. 
Furthermore, we used additional sources, including the NVD dataset, as a complement to the \url{cve.org} dataset, and stored CISA KEV, VulnCheck KEV (KEV-V), Metasploit, Nuclei, Exploit-DB, and CVEs from AlienVault OTX with manual verification~\cite{alienvaultotx} and the ransomware dataset. In the remainder of this paper, we will refer to the CVEs from CISA KEV, KEV-V, and AlienVault OTX as KEVs.

\subsection{Data Collection}\label{sec:data-collection-gh}
All data collection and processing were performed on a Dell laptop with an Intel(R) Core(TM) Ultra 9 185H processor and 32GB of memory. Collecting CVEs from 1999 to 2026 (up to 11 February) took 1 day and 12 hours. A single GitHub access token was used for data collection.

\begin{algorithm}[t]
\caption{Collecting GitHub Exploit Repositories with Depth Partitioning}
\label{alg:github-poc}
\small
\DontPrintSemicolon
\ForEach{$y \in \{1999,\ldots,Y_{\text{current}}\}$}{
      $q \leftarrow$ exact phrase ``CVE-$y$-'' , with qualifier \textsf{fork:false}, sorted by \textsf{updated}\;
      $\mathrm{repos} \leftarrow \textsc{SearchRepositories}(q)$\;
      \ForEach{$r \in \mathrm{repos}$}{
        $\mathrm{ids} \leftarrow \textsc{ExtractCVEsInOrder}(r)$ \tcp*[f]{title $\to$ description $\to$ topics $\to$ README}\;
        \If{$1 \le |\mathrm{ids}| \le 10$}{
          \textsc{Store}\big($r$, metadata, stars history, forks history\big)\;
        }
      }
}
\end{algorithm}
To collect repositories containing exploits, we leveraged the GitHub API. Because the GitHub search API imposes a 1,000 results-per-query cap (10 pages, 100 results per page)~\cite{akhoundali2025eradicating}, when the limit is reached, we recursively limit the query based on repository creation date boundaries. We collect repositories by iterating over each year in the range. To further help stay within rate limits, we did not include forks and instead collected forks of each repository after it was independent from the repository search. This results in Algorithm~\ref{alg:github-poc}, where we use the exact-phrase query \texttt{``CVE-\{YEAR\}-'' fork:false}. This query is informed by the study of malicious GitHub exploits done by El Yadmani et al.~\cite{elyadmani2025securepoc}. 
The search results include repositories where CVE IDs are mentioned in the title, about, tags, or README. We searched for CVE IDs in each section in this order. 

When CVE ID(s) are found, we stop looking in other sections and collect the repository metadata, related CVEs, fork history, and star history. The fork and star history includes timestamps and user IDs. If the repository contains no CVE-ID or more than 10 CVE-IDs, it is skipped. We limit the number of CVEs associated with a repository because it is ambiguous how each CVE contributes to the repository's popularity. By default, when calculating and evaluating the GPM metric, we consider only repositories that mention a single CVE. To provide future researchers with a more comprehensive dataset, we include repositories that mention up to 10 CVEs, but this is only for future work and is not used in our analysis unless explicitly mentioned. This will be further discussed in Section~\ref{sec:gpm}.

To ensure that there are no duplicates, we enforced database-level unique constraints on the repository ID and CVE ID. To enable further processing, we parsed the data into different tables in a Django web app using a PostgreSQL database. This relational data can be queried and presented to the community.
Finally, we note that validating exploits and malicious content screening were out of scope and not part of our data collection, as we primarily use exploits collected from GitHub as a metric to measure attention to a CVE. 

\subsection{Data Analysis}
\noindent \textbf{Summary of the collected data.}
Table~\ref{tbl:collected_data} summarizes the collected data. In total, we collected 332,590 CVEs discovered from June 7, 1999, to February 11, 2026, with all statuses (``REJECTED'' or ``PUBLISHED''), and identified \num{19572} GitHub repositories that mentioned between 1 and 10 ($m$ between 1 and 10) CVEs. Among the CVEs, we collected \num{712374} unique references from CVE descriptions with tags. In GitHub repositories, we collected data on \num{325356} stars and \num{89226} forks, including their timestamps. Out of \num{19572} repositories, \num{19024} (97.20\%) were associated with exactly one CVE (i.e., each repository typically corresponds to a single distinct CVE rather than aggregating multiple CVEs). In total, we collected GitHub exploit repositories for \num{8453} unique CVEs, 
which is approximately 2.5\% of all CVEs (``PUBLISHED'' and ``REJECTED''). We collected 4,610 distinct CVEs for all KEVs combined (CISA KEV, KEV-V, AlienVault OTX), 141 for ransomware CVEs, 971 CVEs in Metasploit modules, 3,867 CVEs for Nuclei, and 132,646 CVEs with SSVC ADP enrichments.

\begin{table}[t]
\centering
\caption{Collected CVE and exploit statistics. \\ $^*$ Reserved/Unknown CVEs refer to CVEs that are mentioned in GitHub, but not found in CVE Program dumps.
}
\label{tbl:collected_data}
\begin{tabular}{lc}
\toprule
\textbf{Type} & \textbf{\# Collected} \\
\midrule
All CVEs  & \num{332590} \\     
Published CVEs & \num{315354} \\
Rejected CVEs & \num{17236} \\
Reserved/Unknown CVEs$^*$ & \num{383} \\
Deferred CVEs (subset of Published) & \num{94588} \\
Unique URL references from CVEs & \num{712374} \\
\midrule
CVEs with SSVC & \num{132646} \\
CISA KEV CVEs & \num{1513} \\
KEV-V CVEs & \num{4599} \\
EPSS CVEs & \num{327711} \\
Ransomware CVEs & \num{141} \\
AlienVaultOTX CVEs & \num{14} \\
All KEVs & \num{4610} \\
\midrule
Metasploit CVEs & \num{971} \\
Nuclei CVEs & \num{3867} \\
\midrule
\multicolumn{2}{l}{\textit{GitHub Repository CVE Exploits}} \\
\midrule
GitHub exploit repositories ($m=1$) & \num{19024} \\
CVEs with exploits in GitHub ($m=1$) & \num{7978} \\
\quad Published & \num{7568} \\
\quad Rejected & \num{51} \\
\quad Reserved/Unknown & \num{359} \\
\midrule
GitHub exploit repositories ($m=10$) & \num{19572} \\
CVEs with exploits in GitHub ($m=10$) & \num{8453} \\
\quad Published & \num{8011} \\
\quad Rejected & \num{59} \\
\quad Reserved/Unknown & \num{383} \\
\midrule
GitHub stars & \num{325356} \\
GitHub forks & \num{89226} \\
\bottomrule
\end{tabular}
\end{table}
\vspace{2pt}

\begin{figure*}[t]
  \centering
  \includegraphics[width=1\textwidth, keepaspectratio]{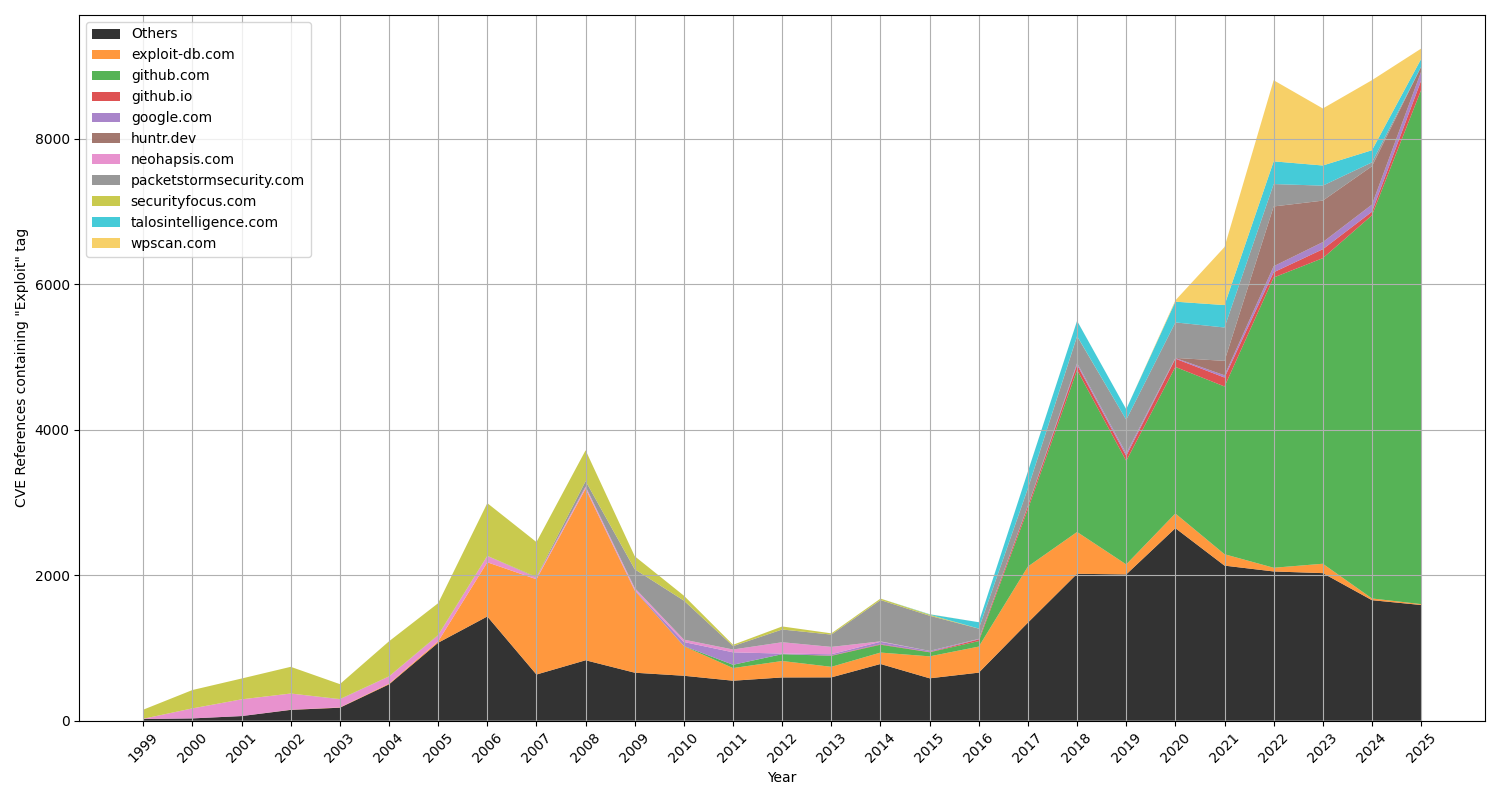}
  \caption{Top domains in CVE references containing the ``exploit'' tag (we do not show 2026 due to partial data).}
   \label{fig:top-site-exploits}
\end{figure*}

\noindent \textbf{Popular exploit sharing platforms.}
From our collected data, we observe that the majority of exploits were published in Exploit-DB between 2006 and 2010, whereas from 2018 to 2025, the majority were published on GitHub. This comprises \num{28330} of \num{59504} references (48\%) with exploit tags for CVEs issued between 2018 and 2025. This clearly indicates a significant shift in trend from Exploit-DB to GitHub. A graphical representation of these trends is shown in Fig.~\ref{fig:top-site-exploits}. 
We suspect a possible reason for this change might be the nature of open-source, where individuals may freely publish their exploits on GitHub, where people can publish their code without delays introduced by peer reviews, and the repository will be shown under their profile. Moreover, Exploit-DB still contains many exploits, but they are not always directly linked to CVE references.


\takeaway{1}{Exploit-DB dominated CVE exploit sharing from 2006–2010, but since 2018, GitHub has been the leading source; 2.5\% of CVEs have a publicly available exploit on GitHub.}

\begin{figure}[t]
  \centering
  \includegraphics[width=0.8\linewidth, keepaspectratio]{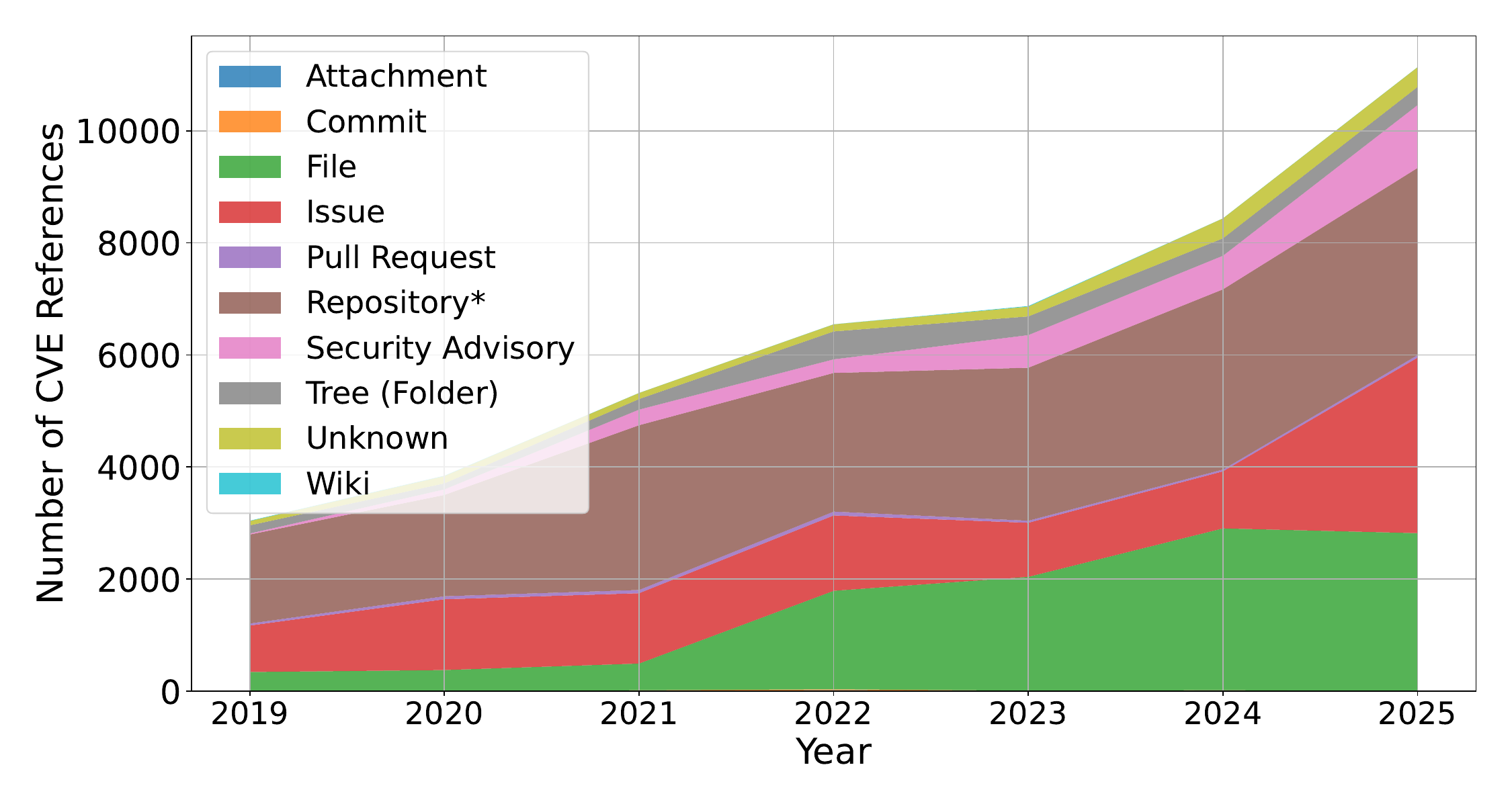}
  \caption{Category of GitHub links found in CVE references.  The``Repository'' type combines mined CVE exploits from GitHub with CVE references tagged as ``exploit''.
  }
   \label{fig:top-exploit-types}
\end{figure}

\noindent \textbf{Exploits on GitHub.}
To understand further where exploits are stored, we analyzed the GitHub link types in CVE references to GitHub repositories. Based on manual analysis, we realized repositories containing CVE exploits are sometimes not listed in CVE references. 
To investigate this, we mined CVE exploits from GitHub repositories not included in the CVE references, as described in Section~\ref{sec:data-collection-gh}. We combined the collected repositories with CVE references to GitHub repositories and de-duplicated them, yielding a complete list of GitHub repositories that contain exploits. The number of found repositories is shown in Fig.~\ref{fig:top-exploit-types} along with a breakdown of the link type of CVE references to other GitHub resources. As can be seen, exploits are shared in three major places. (1) \textbf{Files} in a repository: usually a collection of exploits for multiple CVEs. (2) \textbf{Issues}, as also manually verified, security researchers or developers discuss security issues in the GitHub repository issues section, where everyone can see and participate in the discussion publicly. We note that publicly disclosing security issues before fixing them poses a significant security risk, as attackers gain an exploitation window to access vulnerable systems before a patch is developed~\cite{cipollone2025automating}. (3) \textbf{Repositories}, the largest set, usually contain an exploit for a single CVE. As repositories have popularity-related metrics such as stars and forks, it is possible to map such popularity to CVE exploits.

\begin{figure}[t]
  \centering
\includegraphics[width=0.88\linewidth, keepaspectratio]{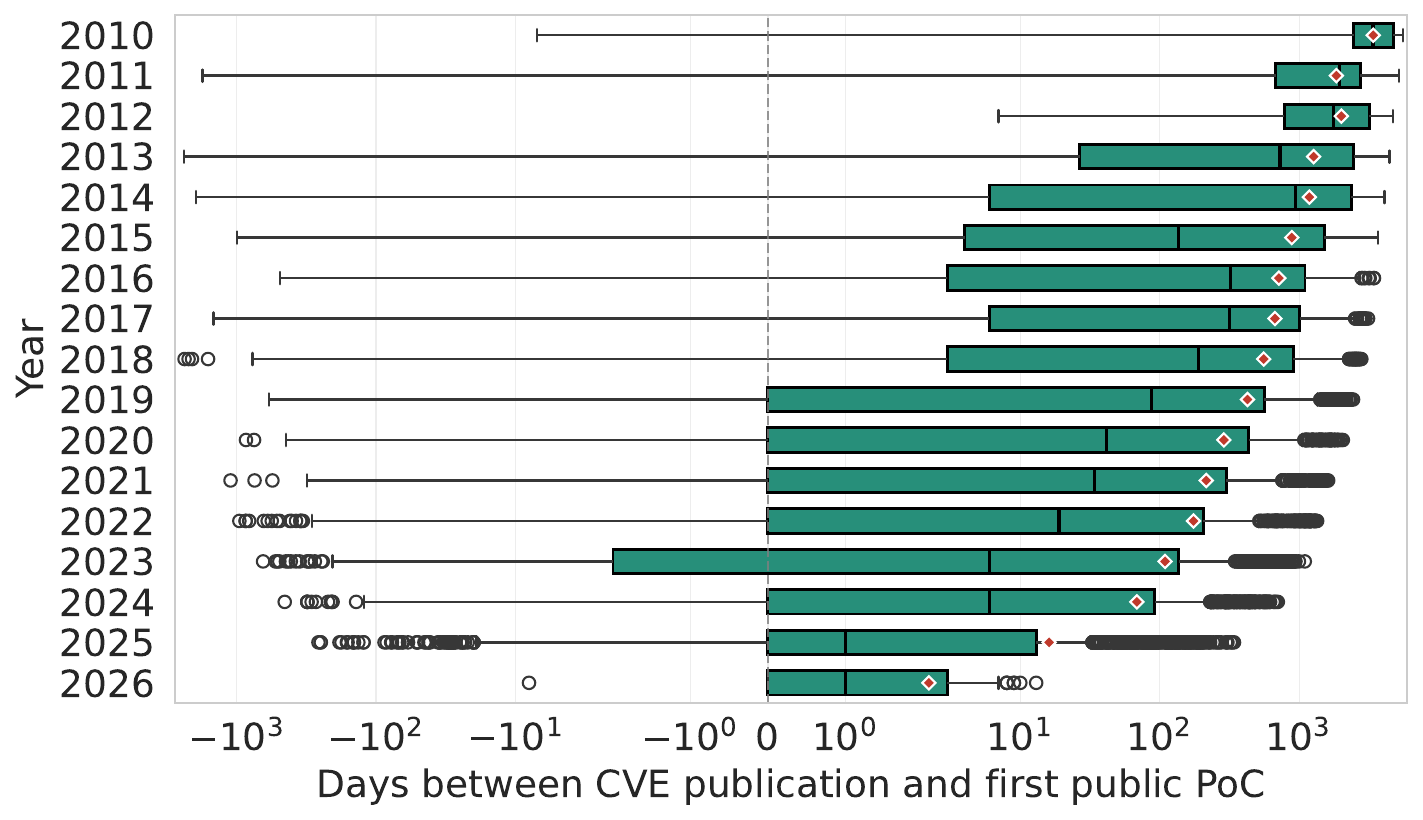}
  \caption{Yearly day gap between CVE publication and exploit release on GitHub.}
   \label{fig:day_diff_cve_publish_vs_poc}
\end{figure}

Furthermore, we investigated the difference between the exploit release date on GitHub and the CVE publication time. The yearly analysis in Fig.~\ref{fig:day_diff_cve_publish_vs_poc} shows a clear decrease in lag between the time of CVE publication and the time the first related GitHub exploit was published on GitHub. In 2023 and 2024, the median lag between exploit publication and CVE publication was 6 days, whereas in 2025, the median is 1 day. In some cases, exploits were available before CVEs were published.

\takeaway{2}{The gap between exploit availability and CVE publication narrows each year.}

\noindent \textbf{CVE repository noise.} We collect GitHub repositories with CVE exploits to measure how much attention CVEs in GitHub receive. However, there may be noise in the collected repositories, as some repositories may contain an incorrectly written CVE-ID (typo) or an original software repository may include a note to a CVE-ID but nothing else. We treat such cases as noise and define valid repositories as those containing any of the following pieces of data: full exploit code, exploit proof-of-concept (PoC), reproduction example, scanner, patcher, repository without functional code, or suspicious\footnote{Some repositories may have obfuscated or compiled code, sometimes with malware~\cite{elyadmani2025securepoc}.} CVE-related repositories. To quantify this noise in the dataset, one of the authors, with more than 5 years in the cybersecurity field, both in academia and industry, manually reviewed 200 repositories. The repositories were selected using stratified random sampling, covering 2018 to 2026, and the selected bins were sampled based on the total number of PoC repositories available in each year. Of the 200 repositories, only 2 contained a CVE typo. Among the remaining 198 repositories at the time of review, the owners of 5 repositories were removed from GitHub (possibly due to malicious code), and 2 repositories were removed, while their owners remained. 3 repositories contained suspicious code, and the rest were valid according to the discussed valid metric, yielding a 1\% noise ratio in the reviewed set. Thus, we can conclude that the data we collect from GitHub largely reflects actual CVE-related content and may be used as an indicator of interest towards particular CVEs. 

\section{The GitHub Popularity Metric (GPM)}\label{sec:gpm}

To measure the popularity of CVEs, we collected CVE-related repositories on GitHub, along with metadata on relevant user and engagement events, such as star and fork counts.

\subsection{The GPM Metric Definition}
We measure the popularity of a CVE $c$ by counting the number of unique user IDs that interacted with a CVE exploit repository via repository creation, starring, or forking.  
Each user contributes at most $1$ to $\mathrm{GPM}(c)$, regardless of how many interactions or repositories (linked to $c$) they engage with. If a repository is associated with multiple CVEs (at most $m$), a single interaction contributes $1$ to each CVE. Unless explicitly stated otherwise, we only considered repositories with $m=1$. 
To reduce false-positive mappings, we included only repositories created after the corresponding CVE reservation date. 

\noindent \textbf{Formal GPM definition.}
Let $\mathcal{U}$ be the set of GitHub users, $\mathcal{R}$ the set of repositories, and $\mathcal{C}$ the set of CVEs.
Let \linebreak $\mathcal{K} = \{\textsc{Create}, \textsc{Star}, \textsc{Fork}\}$ denote the interaction types, and let  
$\mathcal{E} \subseteq \mathcal{U} \times \mathcal{K} \times \mathcal{R}$ be the collected interactions set.
Let $\mathcal{A} \subseteq \mathcal{R} \times \mathcal{C}$ be the repository–CVE association relation.  
Let $m \in \{1, 2, 3, \dots\}$ be the maximum number of CVEs a repository may be associated with.

\begin{equation*}
\forall r \in \mathcal{R}:\quad
\left| \left\{ c \in \mathcal{C} : (r, c) \in \mathcal{A} \right\} \right| \le m.
\label{eq:max-cves}
\end{equation*}

We define the popularity of CVE $c\in\mathcal{C}$ as the number of \emph{distinct user identifiers from $\mathcal{E}$} who interacted with at least one repository associated with $c$: 
\begin{equation}
\mathrm{GPM}(c)
\;=\;
\Bigl|\bigl\{\,u\in\mathcal{U}\; \big|\; \exists\, r\in\mathcal{R},\, k\in\mathcal{K}:
(r,c)\in\mathcal{A}\ \wedge\ (u,k,r)\in\mathcal{E}\,\bigr\}\Bigr|.
\label{eq:gpm}
\end{equation}

\subsection{Evaluation Metrics}
We evaluate the effectiveness of the GPM using CISA KEV as ground truth. First, we create an initial set of CVEs, denoted by $C$, present in GPM, and compare their exploitation status against the KEV list. 
We treat the GPM thresholding as a binary classifier. A CVE is considered a ground-truth positive if it is confirmed as exploited (i.e., listed in KEV), and a ground-truth negative if its popularity (the GPM value) is higher than the set threshold but it is not in KEV.

Let $\mathcal{C}^+ \subseteq \mathcal{C}$ denote the set of ground-truth positives, and let
\begin{equation}
{\mathcal{C}}^+(\tau)
\;=\;
\{\, c \in \mathcal{C} \mid \mathrm{GPM}(c) \ge \tau \,\}
\label{eq:predicted}
\end{equation}
be the set of predicted positives at the popularity threshold $\tau$.

\noindent We then define the standard metrics:

\noindent \textbf{Precision.}
The fraction of predicted positives that are correct:
\begin{equation}
\mathrm{Precision}(\tau)
\;=\;
\frac{|{\mathcal{C}}^+(\tau) \cap \mathcal{C}^+|}
     {|{\mathcal{C}}^+(\tau)|}.
\label{eq:precision}
\end{equation}

\noindent \textbf{Recall.}
The fraction of ground-truth positives that are recovered:
\begin{equation}
\mathrm{Recall}(\tau)
\;=\;
\frac{|{\mathcal{C}}^+(\tau) \cap \mathcal{C}^+|}
     {|\mathcal{C}^+|}.
\label{eq:recall}
\end{equation}

\noindent \textbf{F1-score.}
The harmonic mean of precision and recall:
\begin{equation}
\mathrm{F1}(\tau)
\;=\;
\frac{2 \cdot \mathrm{Precision}(\tau)\cdot \mathrm{Recall}(\tau)}
     {\mathrm{Precision}(\tau)+\mathrm{Recall}(\tau)}.
\label{eq:f1}
\end{equation}

\begin{figure}[t]
  \centering
  \includegraphics[width=0.8\linewidth, keepaspectratio]{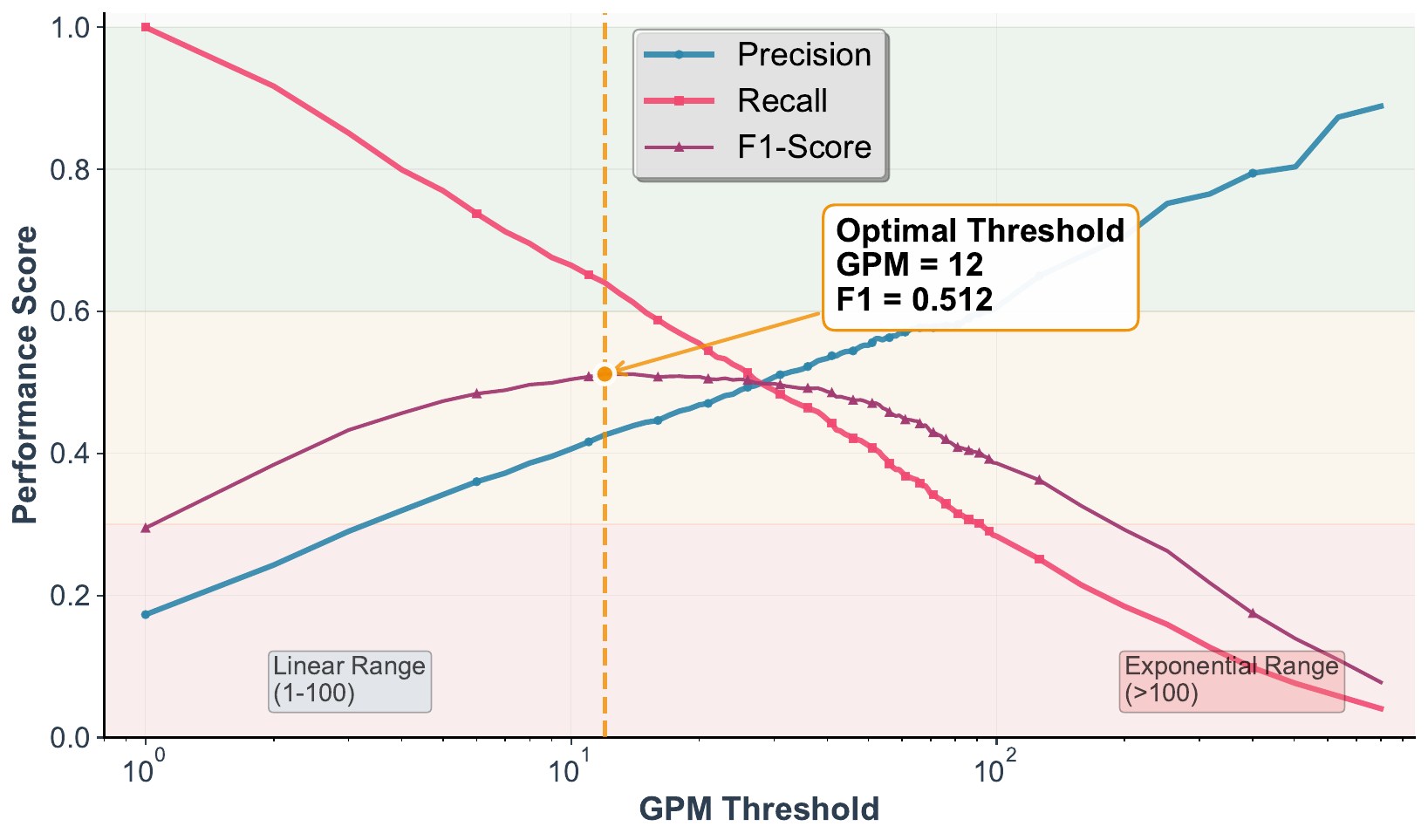}
  \caption{The GPM threshold performance compared with exploited CVEs.}
   \label{fig:f1score-threshold}
\end{figure}

As shown in Fig.~\ref{fig:f1score-threshold}, varying the threshold $\tau$ yields a trade-off: higher thresholds increase precision (CVEs with high GPM scores are more likely to be exploited) but decrease recall (coverage will be limited). The optimal threshold $\tau^\star$ is selected to maximize $\mathrm{F1}(\tau)$, balancing precision and recall. To better understand and evaluate the predictive effectiveness for an optimal GPM threshold, we first compute the GPM threshold and F1 score for a specific year (as a training year) and then use the obtained optimal threshold ($\tau_{tr}$) to evaluate its effectiveness on the next year's data, measured as $\mathrm{F1}(\tau_{tr})$ on that subsequent year. Then we calculate the optimal threshold again by including the test year ($\tau_{te}$) and calculate the F1 score using this threshold. Finally, we report both the absolute F1 difference, 
$\Delta \mathrm{F1} = \mathrm{F1}(\tau_{tr}) - \mathrm{F1}(\tau_{te})$, 
and the relative F1 difference, 
$\frac{\mathrm{F1}(\tau_{tr}) - \mathrm{F1}(\tau_{te})}{\mathrm{F1}(\tau_{te})}$. The results are shown in Table~\ref{tab:gpm_threshold_generalization}. Although the optimal threshold between 2020 and 2026 ranged from 12 to 27, using the historical threshold yielded only a consistently small performance drop. Thus, despite the relatively wide range of optimal thresholds, F1 performance remains robust. Based on the latest available data, a GPM threshold of 12 yields the best F1 score, balancing precision and recall. 

\begin{table}[t]
\centering
\caption{Threshold Generalization: Optimal $\tau$ trained on Year $t$ ($\tau_{tr}$) and evaluated on Year $t{+}1$. Theoretical maximum F1 for the test year is denoted by F1($\tau_{te}$).}
\label{tab:gpm_threshold_generalization}
\begin{tabular}{cccccccc} 
\toprule
\textbf{Yr$_{tr}$} & \textbf{Yr$_{te}$} & $\boldsymbol{\tau_{tr}}$ & $\boldsymbol{\tau_{te}}$ & \textbf{F1 ($\tau_{tr}$)} & \textbf{F1 ($\tau_{te}$)} & $\boldsymbol{\Delta}$\textbf{F1} & \textbf{Rel. $\boldsymbol{\Delta}$F1} \\
\midrule
2020 & 2021 & 27 & 11 & 0.508 & 0.522 & -0.014 & -0.027 \\
2021 & 2022 & 11 & 25 & 0.505 & 0.519 & -0.014 & -0.028 \\
2022 & 2023 & 25 & 22 & 0.514 & 0.518 & -0.004 & -0.008 \\
2023 & 2024 & 22 & 17 & 0.513 & 0.519 & -0.006 & -0.011 \\
2024 & 2025 & 17 & 13 & 0.509 & 0.513 & -0.004 & -0.008 \\
2025 & 2026 & 13 & 12 & 0.511 & 0.512 & -0.001 & -0.002 \\
\bottomrule
\end{tabular}
\end{table}



\begin{figure}[t]
  \centering
\includegraphics[width=0.62\linewidth, keepaspectratio]{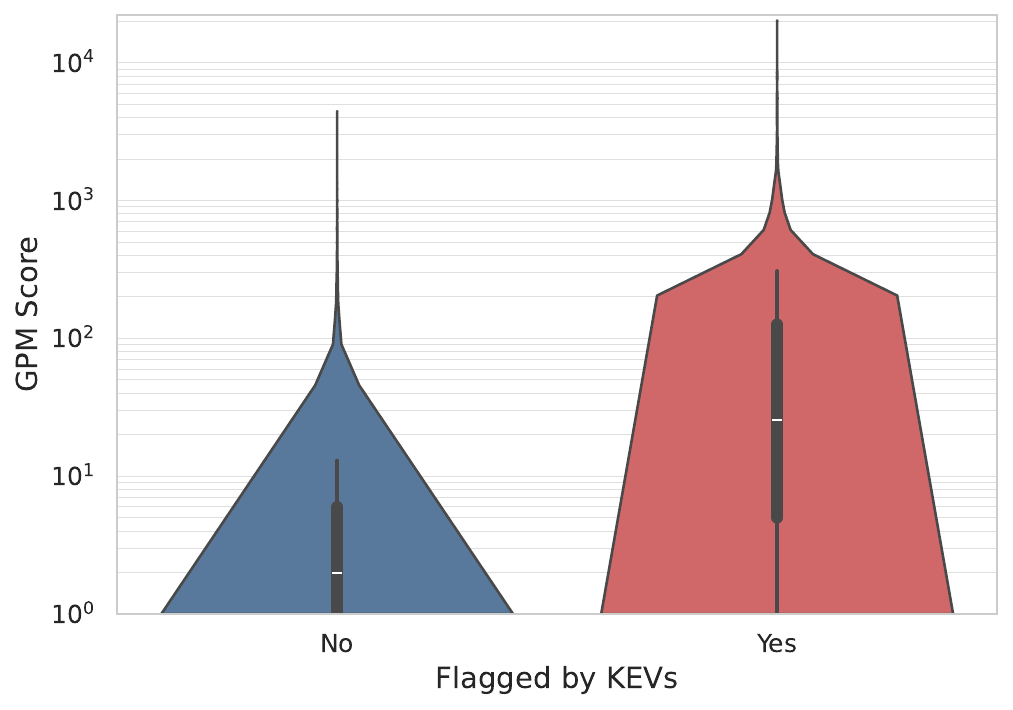}
  \caption{The GPM difference for CVEs with exploit evidence versus other CVEs (log scale).}
   \label{fig:kevs-vs-gpm-flags}
\end{figure}

\subsection{Exploit Popularity vs Addition to KEV}
Fig.~\ref{fig:kevs-vs-gpm-flags} depicts the GPM scores for CVEs that appear in KEVs (and thus are known to be exploited in the wild), compared to CVEs that were not flagged. As can be seen, the exploited CVEs have a noticeably higher GPM score. It may be that, in addition to a KEV list, other factors influence CVE popularity, causing the GPM to increase more rapidly. To rule out such post-treatment bias, we used historical GPM data and measured interactions of each CVE before and after the data was added to the CISA KEV (for which the addition date is known).

\noindent \textbf{Metric 1: Pre/Post symmetric slopes.}
For each CVE $c$ with the CISA KEV publication date $\text{KEV}(c)$, we calculate a list of daily interactions around the KEV inclusion date. Let $y_t^{(c)}$ be the number of observed interactions (the GPM) with the repositories for this CVE on day offset $t \in \mathbb{Z}$, with $t=0$ corresponding to the KEV publication date itself. Offsets are defined as
\[
t = \big(\texttt{date} - \text{KEV}(c)\big)\ \text{in days},
\]
offset days without interactions are assigned $y_t^{(c)}=0$.

Given a symmetric days window width $W$, we compute two fixed-length sequences of GPM counts, excluding the KEV day:
\begin{align*}
\mathbf{y}^{(c)}_{\mathrm{pre}}(W) &= \big( y_{-W}^{(c)}, y_{-(W-1)}^{(c)}, \ldots, y_{-1}^{(c)} \big) \qquad \\
\mathbf{y}^{(c)}_{\mathrm{post}}(W) &= \big( y_{+1}^{(c)}, y_{+2}^{(c)}, \ldots, y_{+W}^{(c)} \big) \qquad
\end{align*}

For each sequence, we computed a least-squares regression slope over the $W$ days. Accordingly, for $\mathbf{y} = (y_1,\dots,y_W)$ indexed by $x=1,\dots,W$, we fit
\[
\hat{\beta}_1 = \arg\min_{\beta_0,\beta_1} \sum_{x=1}^{W} \big(y_x - (\beta_0 + \beta_1 x)\big)^2,
\]
and define
\[
S^{(c)}_{\mathrm{pre}}(W) = \hat{\beta}_1\big(\mathbf{y}^{(c)}_{\mathrm{pre}}(W)\big),
\qquad
S^{(c)}_{\mathrm{post}}(W) = \hat{\beta}_1\big(\mathbf{y}^{(c)}_{\mathrm{post}}(W)\big).
\]

To ensure a well-defined pre-KEV time series, we restricted the analysis to CVEs with at least one recorded GPM value within $W$ days before their KEV inclusion date. We then compare the differences between $S_{\mathrm{pre}}(W)$ and $S_{\mathrm{post}}(W)$ in CVEs using a paired Wilcoxon signed-rank test.

\noindent \textbf{Metric 2: Pre/post maximum GPM ``W''-days.}
For every CVE $c$ with CISA KEV date $\mathrm{KEV}(c)$, we create a series of daily interactions $y_d^{(c)}$. We split this series into pre-KEV ($d < \mathrm{KEV}(c)$) and post-KEV ($d > \mathrm{KEV}(c)$) segments, ignoring the KEV day.
For a window size of $W$ days, we compute rolling-window sums

\[
R_i^{(c)}(W) = \sum_{j=i}^{i+W-1} y_j^{(c)},
\]
The maximum rolling sum computes the burst size:

\begin{align*}
M^{(c)}_{\mathrm{pre}}(W) = \max_{i: d_i < \mathrm{KEV}(c)} R_i^{(c)}(W) \qquad \\
M^{(c)}_{\mathrm{post}}(W) = \max_{i: d_i > \mathrm{KEV}(c)} R_i^{(c)}(W) \qquad 
\end{align*} 
Then, for each Pre/Post burst size, we test $(M^{(c)}_{\mathrm{pre}}(W), M^{(c)}_{\mathrm{post}}(W))$ between CVEs with a paired Wilcoxon signed-rank test.

\begin{table}[t]
\centering
\caption{Pre vs Post-KEV GPM Slopes}
\label{tbl:prepos-kev-slopes}
\begin{tabular}{ccccccc}
\toprule
\textbf{$W$} & \textbf{CVEs} & \textbf{$\mu$Pre} & \textbf{$\mu$Post} & \textbf{$p$} & \textbf{$\Delta$} & \textbf{Sig.} \\
\midrule
2 & 493 & 0.387 & -0.327 & $0.055$ & $\rightarrow$ & No \\
7 & 461 & -0.073 & -0.038 & $0.644$ & $\rightarrow$ & No \\
14 & 443 & -0.009 & -0.024 & $5.13 \times 10^{-4}$ & $\downarrow$ & \textbf{Yes} \\
30 & 412 & -0.001 & -0.008 & $2.05 \times 10^{-4}$ & $\downarrow$ & \textbf{Yes} \\
60 & 382 & -0.003 & -0.003 & $0.135$ & $\rightarrow$ & No \\
90 & 361 & -0.001 & -0.001 & $0.418$ & $\rightarrow$ & No \\
180 & 317 & -0.000 & -0.000 & $0.121$ & $\rightarrow$ & No \\
365 & 259 & -0.000 & -0.000 & $0.913$ & $\rightarrow$ & No \\
\bottomrule
\end{tabular}
\end{table}

\begin{table}[t]
\centering
\caption{Pre/Post Maximum GPM $W$-Days}
\label{tbl:prepos-kev-maxgpm}
\begin{tabular}{ccccccc}
\toprule
\textbf{$W$} & \textbf{CVEs} & \textbf{$\mu$Pre} & \textbf{$\mu$Post} & \textbf{$p$} & \textbf{$\Delta$} & \textbf{Sig.} \\
\midrule
30 & 412 & 175.803 & 13.680 & $3.11 \times 10^{-56}$ & $\downarrow$ & \textbf{Yes} \\
60 & 382 & 190.484 & 16.796 & $7.69 \times 10^{-55}$ & $\downarrow$ & \textbf{Yes} \\
90 & 361 & 200.795 & 17.382 & $4.32 \times 10^{-52}$ & $\downarrow$ & \textbf{Yes} \\
180 & 317 & 211.442 & 19.782 & $4.02 \times 10^{-46}$ & $\downarrow$ & \textbf{Yes} \\
365 & 259 & 257.741 & 27.571 & $1.20 \times 10^{-38}$ & $\downarrow$ & \textbf{Yes} \\
\bottomrule
\end{tabular}
\end{table}

\vspace{2pt}
\noindent \textbf{Results.}
As shown in Table~\ref{tbl:prepos-kev-slopes}, the popularity slope does not change noticeably after a CVE is added to the KEV list. Moreover, in Table~\ref{tbl:prepos-kev-slopes}, the 14 and 30 $w$ have $p<0.05$ (indicating a statistically significant result), but still the mean-post is negative. Thus, there is no sign of an increase in popularity after the addition to KEV in any case. In the second metric (maximum interaction), as shown in Table~\ref{tbl:prepos-kev-maxgpm}, the direction of the slope change is always downward; thus, CVE popularity actually decreased after it was added to KEV. This also aligns with the first metric. Therefore, we conclude that \textbf{the GPM is predictive rather than reactive} and can be used independently as a sign to detect exploited CVEs, such as those in KEV, without post-treatment bias.

\takeaway{3}{Exploit repositories for known exploited CVEs on GitHub tend to be more popular. This popularity is not biased by the addition to the CISA KEV list.}


\section{Coverage and Insights from thw GPM}\label{sec:insightsGPM}
\noindent \textbf{Coverage.}
As described in Section~\ref{sec:data}, as of 2026-02-11, the number of CVEs with a ``PUBLISHED'' status (excluding ``REJECTED'' and unknown CVE statuses) that are covered by our dataset is \num{7568} CVEs with GPM $m=1$ and \num{8011} CVEs with GPM $m=10$. This is equivalent to 2.39\% (GPM $m=1$) and 2.54\% (GPM $m=10$) of all published CVEs (\num{315354} in total).

\begin{figure}[t] 
  \centering
\includegraphics[width=0.52\linewidth, keepaspectratio]{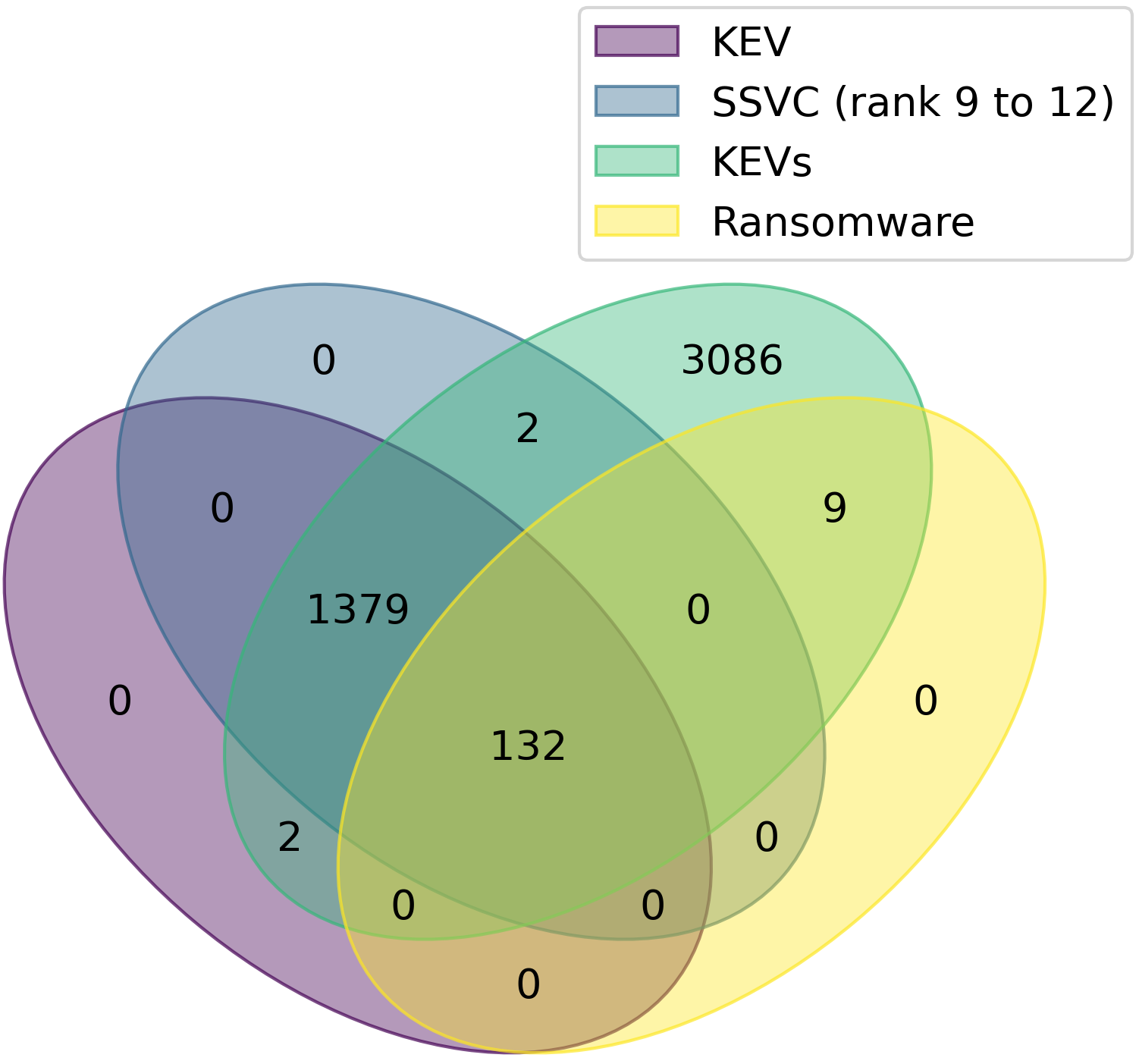}
  \caption{Venn diagram of CVEs in CISA KEV, KEVs (VulnCheck + AlienVaultOTX), actively exploited SSVC CVEs, and the ransomware dataset.}
   \label{fig:venn-ssvc-gh-epss-gh}
\end{figure}

\noindent \textbf{CVEs with exploits solely covered by the GPM.}
To assess whether GPM can identify popular CVEs with available exploits that are not present in other KEVs, we compare GPM's coverage with that of other KEVs.
Fig.~\ref{fig:venn-ssvc-gh-epss-gh} summarizes the number of CVEs present in different KEVs.
As can be seen, despite the diversity of the sets, CISA KEV, VulnCheck KEV, and our Alienvault OTX set contain the larger set. Because KEVs is the superset, we use the KEVs set as the ground truth set of CVEs for analysis.

\begin{figure}[t] 
  \centering
\includegraphics[width=0.6\linewidth, keepaspectratio]{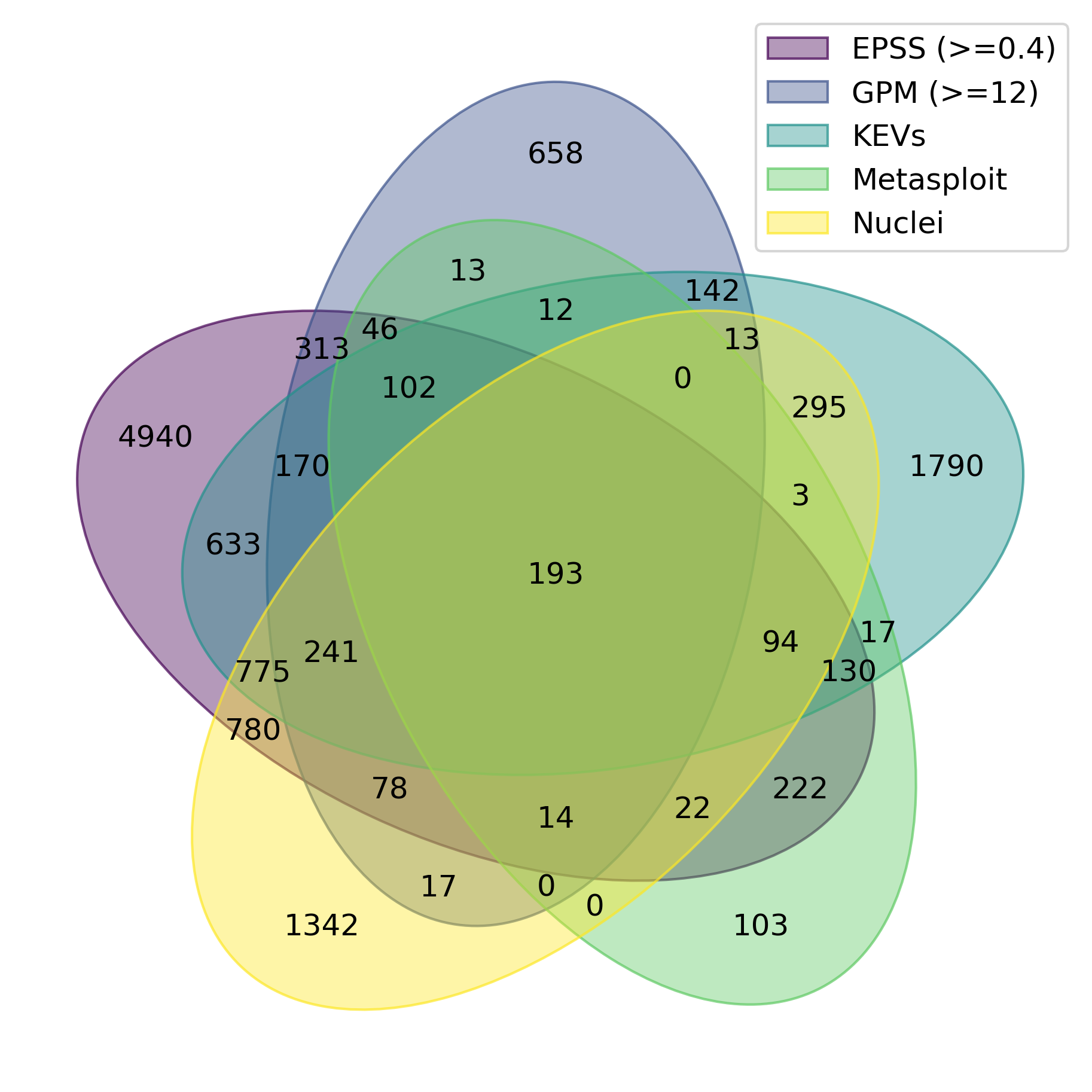}
  \caption{Venn diagram of CVEs in KEVs, the EPSS ($\geq$0.4), Nuclei templates, Metasploit modules, and the GPM ($\geq$12).}
   \label{fig:venn5-kevs-ghm}
\end{figure}

Next, we compare the overlap of CVEs across KEVs, exploit sources (Metasploit, Nuclei, GitHub), and the EPSS, as shown in Fig.~\ref{fig:venn5-kevs-ghm}. The statistics of each set, compared with GPM ($\geq$12, the threshold determined in Section~\ref{sec:gpm}) and GPM (all samples), are shown in Table~\ref{tab:cve_overlap_unique}. The first key observation is that only 193 CVEs appear across all sources, indicating that \textbf{no single source is sufficient to cover all CVEs}. Each source also contributes a substantial number of unique CVEs, so relying on any one source provides only a partial perspective of the exploitation landscape. In particular, the GPM identifies 658 CVEs with a GPM score of 12 or higher that are absent from any other dataset. This suggests that GitHub contains unique exploits that warrant further analysis and that leveraging the GPM, either independently or in combination with other sources, can provide valuable additional coverage.
\begin{table}[t]
\centering
\setlength{\tabcolsep}{4pt}
\caption{CVE overlap counts per data source vs GPMs}
\label{tab:cve_overlap_unique}
\scalebox{1}{
\begin{tabular}{lccc}
\toprule
\textbf{GPM (>=12)} & Intersection & Unique to ref & Unique to source \\
\midrule
EPSS (>=0.4) & \num{1157} & 855 & \num{7596} \\
KEVs & 873 & \num{1139} & \num{3737} \\
Metasploit & \num{380} & \num{1632} & 591 \\
Nuclei & 556 & \num{1456} & \num{3311} \\
Ransomware & 91 & \num{1921} & 50 \\
\toprule
\textbf{GPM (all)} & Intersection & Unique to ref & Unique to source \\
\midrule
EPSS (>=0.4) & \num{2212} & \num{5766} & \num{6541} \\
KEVs & \num{1398} & \num{6580} & \num{3212} \\
Metasploit & 510 & \num{7468} & 461 \\
Nuclei & \num{1081} & \num{6897} & \num{2786} \\
Ransomware & 100 & \num{7878} & 41 \\
\bottomrule
\end{tabular}
}
\end{table}

\vspace{2pt}
\noindent \textbf{Comparison with the CVSS.}
To compare CVSS base scores with the GPM, we examine differences in the distribution of the CVSS across the \emph{GPM set} (where a CVE has at least one GitHub repository) and the \emph{non-GPM set}. We selected CVEs in recent years (2018 up to 2026), and partitioned scores into left-closed, right-open ranges\footnote{Except $[9,10]$,
which is closed on both ends to include the maximum score of 10}:
\[
\mathcal{R}=\{[0,4),[4,7),[7,9),[9,10]\},
\]
For each range \(R\in\mathcal{R}\), let
\(n_{\mathrm{GPM}}(R)\) and \(n_{\mathrm{non\text{-}GPM}}(R)\) be the number of CVEs
whose CVSS base score falls in \(R\), and let \(N_{\mathrm{GPM}}\) and
\(N_{\mathrm{non\text{-}GPM}}\) be the total number of CVEs in each set. We compute the within-set
prevalence of \(R\) as
\begin{align*}
p_{\mathrm{GPM}}(R) &= 100 \times \frac{n_{\mathrm{GPM}}(R)}{N_{\mathrm{GPM}}},\\
p_{\mathrm{non\text{-}GPM}}(R) &= 100 \times \frac{n_{\mathrm{non\text{-}GPM}}(R)}{N_{\mathrm{non\text{-}GPM}}}.
\end{align*}
We define the \emph{range-level selection bias} as the percentage-point difference
\begin{equation*}
\Delta(R) = p_{\mathrm{GPM}}(R) - p_{\mathrm{non\text{-}GPM}}(R).
\end{equation*}
Here, \(\Delta(R)>0\) indicates that range \(R\) is over-represented in the GPM set
relative to the non-GPM set, while \(\Delta(R)<0\) indicates under-representation.

\begin{figure}[t]
  \centering
  \includegraphics[width=0.62\linewidth, keepaspectratio]{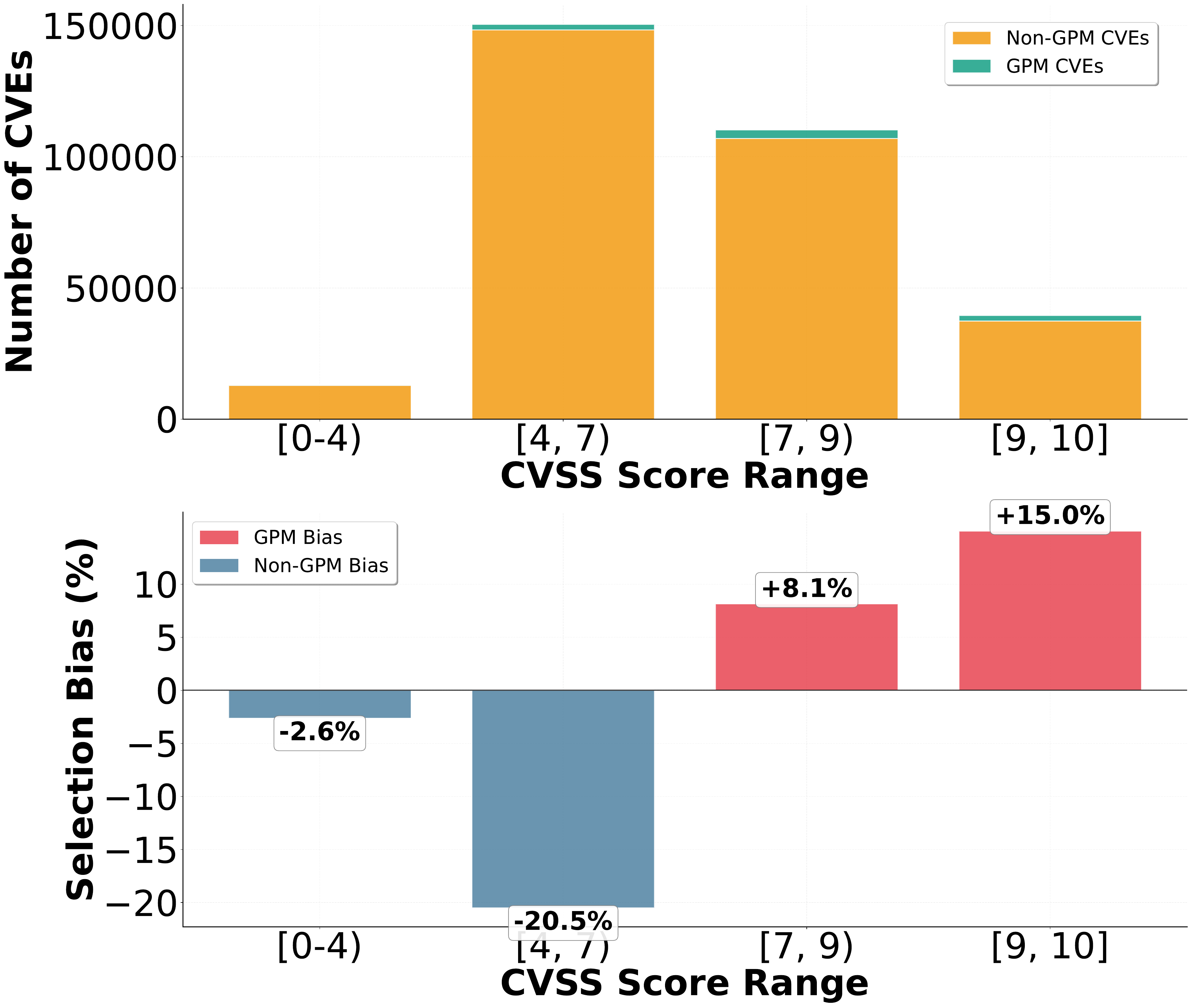}
  \caption{Comparison of CVEs in GPM set versus non-GPM set.}
   \label{fig:cvss-vs-gpm-bias}
\end{figure}

Furthermore, we test whether the GPM and non-GPM sets have different distributions within the defined CVSS score ranges based on severity using a Pearson chi-square test of independence (4 bins × 2 groups). The effect size was quantified using Cramér’s $V$ (0 = no association, 1 = perfect association).
As shown in Fig.~\ref{fig:cvss-vs-gpm-bias}, CVSS ranges for the low and medium severity categories (CVSS < 7.0) are biased toward the non-GPM set, whereas ranges at the high and critical severity categories
(CVSS $\geq$ 7.0) are biased toward the GPM set, with the strongest over-representation in the critical severity category CVSS (9.0–10.0). This indicates that CVEs in the GPM set are predominantly high-severity. The Pearson chi-square test also indicates a statistically significant association, with $\chi^2(3)=2249.93$ and $p < 0.001$\footnote{We consider results significant if $p < 0.05$.}, and a small effect size (Cramér's $V=0.103$). Counts for the different CVSS ranges, including the GPM set rate, are shown in Table~\ref{tab:cvss-by-group}. The GPM rate increases monotonically with the CVSS score range.

\begin{table}[t]
\centering
\caption{CVSS distribution by group (counts and within-group rates).}
\label{tab:cvss-by-group}
\scalebox{1}{
\begin{tabularx}{0.58\columnwidth}{Xrrrr}
\toprule
\textbf{Range} & \textbf{GPM} & \textbf{Non-GPM} & \textbf{Total} & \textbf{GPM Rate} \\
\midrule
$[0,4)$ & 99 & \num{6807} & \num{6906} & 1.4\% \\
$[4,7)$ & \num{1843} & \num{99251} & \num{101094} & 1.8\% \\
$[7,9)$ & \num{2839} & \num{75159} & \num{77998} & 3.6\% \\
$[9,10]$ & \num{1837} & \num{22550} & \num{24387} & 7.5\% \\
\midrule
\textbf{Total} & \textbf{\num{6618}} & \textbf{\num{203767}} & \textbf{\num{210385}} & \textbf{3.1\%} \\
\bottomrule
\end{tabularx}
}
\end{table}

\vspace{2pt}
\noindent \textbf{Comparison with the EPSS.}
Previously, we used the EPSS in a binary fashion (whether the score is higher than a given threshold). However, the EPSS is a numerical metric, and there is no universally correct cut-off value. In general, transforming EPSS scores into bins is discouraged~\cite{RomanoskyJacobs2022EPSSPercentiles}. Thus, EPSS values are better treated as probabilities. Moreover, it is encouraged to use the EPSS when other metrics, such as signs of active exploitation, are not present. Smaller companies may use thresholds, but larger companies are encouraged to use their capacity.  
As both the EPSS and GPM are temporal data, we can compare their behaviour longitudinally. To this end, we used EPSS versions 3 (with most samples) and 4 (the latest version at the time of analysis but with fewer samples) and compared them against the available GPM values. We compute the Spearman rank correlation coefficient $\rho$ for each CVE. 
Given two paired sequences $(x_i, y_i)$ of length $n$, Spearman’s $\rho$ is defined as:
\begin{equation*}
\rho = 1 - \frac{6 \sum_{i=1}^n d_i^2}{n(n^2 - 1)}
\end{equation*}
For each CVE, we align the daily EPSS probability scores $x_i$ with the cumulative counts of GPM interactions $y_i$ (stars, forks, and repository creation events) through day $i$. Pairs where either variable is constant (zero variance) are excluded, as correlation is undefined. Each correlation is accompanied by a $p$-value testing the null hypothesis 
$H_0: \rho = 0$ (no monotonic association).

\begin{table}[t]
  \centering
  \caption{The Spearman correlation of the EPSS versus the GPM; significance evaluated at $p<0.05$.}
  \label{tab:epss-gpm-corr-vert}
    \begin{tabularx}{0.72\linewidth}{l>{\centering\arraybackslash}X>{\centering\arraybackslash}X}
    \toprule
    & \textbf{EPSS V3} & \textbf{EPSS V4} \\
    \midrule
    Start date  & 2023-03-09 & 2025-03-17 \\
    End date    & 2025-03-16 & 2026-02-11 \\
    Samples     & 4,128 & 3,483 \\
    Valid       & 1,702 (41.2\%) & 1,392 (40.0\%) \\
    Mean        & 0.2423 & 0.2274 \\
    Median      & 0.4059 & 0.3270 \\
    Significant & 1621/1702 (95.2\%) & 1244/1392 (89.4\%) \\
    \bottomrule
  \end{tabularx}
\end{table}

As shown in Table~\ref{tab:epss-gpm-corr-vert}, GPM is significantly correlated with EPSS version 3 in 95.2\% of cases and with EPSS version 4 in 89.4\% of cases. Although EPSS version 4 contains fewer samples due to its recent release, the correlation remains high. The experiment shows that the GPM strongly aligns with the EPSS. One limitation is that correlation is undefined when a variable is constant. We also removed cases when EPSS or GPM data is too sparse for a rank correlation (fewer than three EPSS points, or fewer than three distinct GPM interaction days); these cases together account for \textasciitilde{}60\% of samples.

\vspace{2pt}
\noindent \textbf{Comparison with the SSVC.}
To compare the SSVC with the GPM, we used the intersection of CVEs that had both a GitHub repository as an exploit and SSVC values. That is, respectively, 132,646 samples from the SSVC and 7,978 CVEs from GitHub repositories, with a final intersection of 4,329 CVEs in both SSVC and GPM. 

Unlike the SSVC, CVE itself is vendor-neutral; thus, the Mission and Well-Being features were not used in the comparison, and the final decision tree has been flattened and sorted from most important to least important, using generally available data (Technical Impact, Automatability, and Exploitability).\footnote{Decisions were encoded as ordinal values (1–4) and results were ranked by the aggregate decision score.} For a fair comparison, we sorted CVEs by their mean interaction count (the GPM score) and compared the resulting GPM ranking with the SSVC ranking. 

\begin{figure*}[t]
  \centering
  \includegraphics[width=1\textwidth, keepaspectratio]{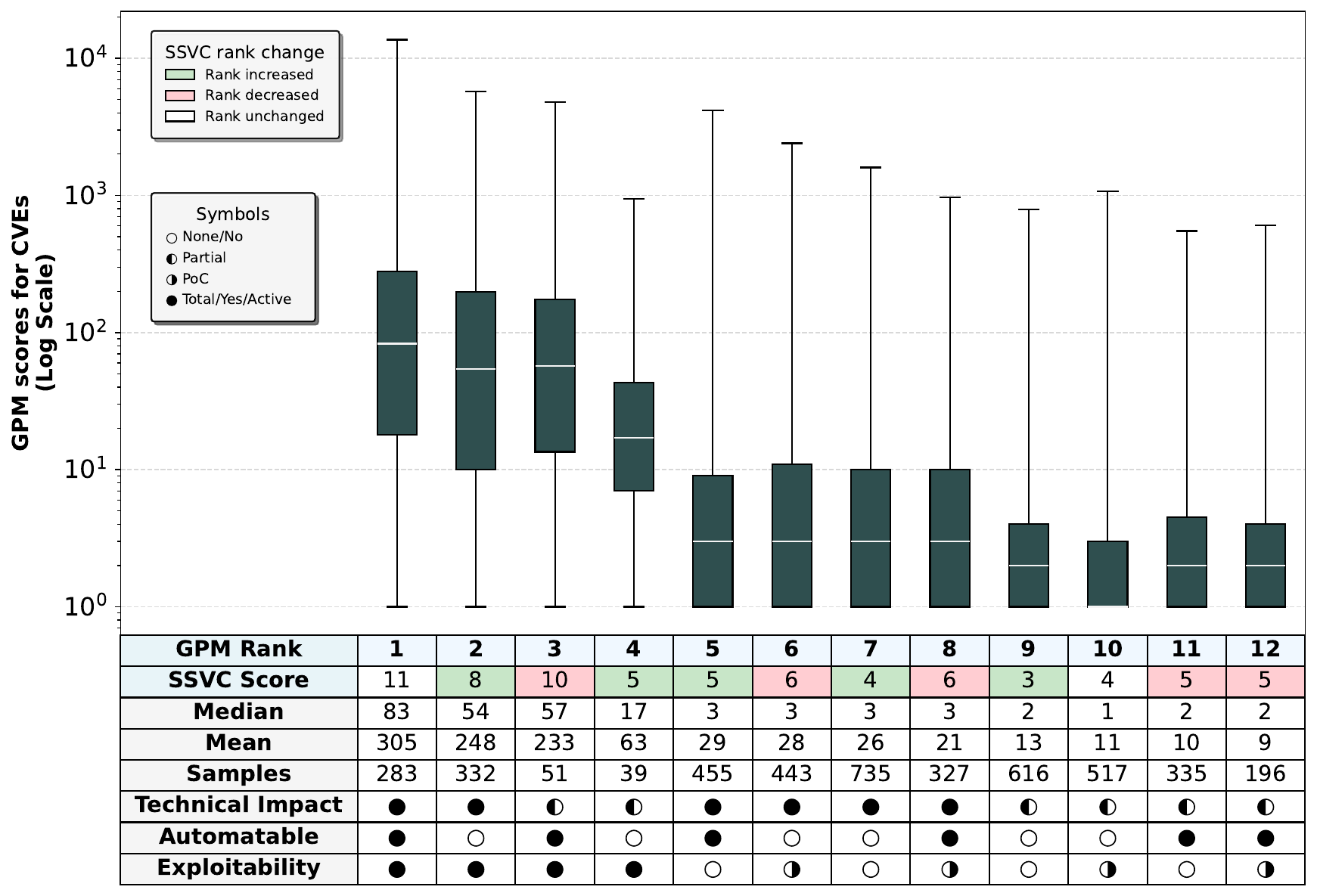}
  \caption{Comparison of SSVC CVE ranks versus the GPM-based ranking, sorted by the GPM mean.}
   \label{fig:ssvc-vs-github-sorted}
\end{figure*}

As can be seen in Fig.~\ref{fig:ssvc-vs-github-sorted}, ``Active'' Exploitability per the SSVC (i.e., the CVE is verified as exploited) is the first decision point for CVEs with the highest GPM mean. This forms the top 4 bracket from rank 1 to 4. In this bracket, the technical impact ``Total'' dominates ranks 1st and 2nd, and ``Partial'' technical impact uses the second bracket. In the 5th to 8th positions, the "Exploitability" field for CVEs is either ``PoC'' or ``None''. The SSVC only collects PoCs from well-known sources such as Exploit-DB, not GitHub repositories. As a result, the ``None'' exploit by SSVC is not necessarily correct and may actually mean a PoC is available in other sources or privately. Thus, when ``Exploitability" is ignored, interestingly, all CVEs within ranks 5th to 8th have a technical impact of ``Total'', which is aligned with the previous finding.

To summarize, SSVC parameters have a meaningful relation when compared to the GPM ranking. Exploitability is the most important factor, followed by Technical impact, which serves as a tie-breaker. The Automation field in SSVC did not show a meaningful correlation.

\takeaway{4.1}{The GPM provides insights into CVEs that are not highlighted by other established metrics. It strongly aligns with the industry-standard EPSS and CVSS metrics and improves the SSVC prioritization.}

\section{Case Studies}\label{sec:case_studies}

\noindent \textbf{Identifying a wrongly rejected CVE.} Early in this study, we realized that a rejected CVE had a high GPM value. This CVE (\textbf{CVE-2024-21683}) had the highest score among CVEs with status ``REJECTED'', at 166. Only a few days after the analysis, the Black Basta ransomware group chat was leaked, and this was one of the CVEs mentioned in their chats. Once the cybersecurity community noticed this, on 15 March 2025, the status of this CVE was changed from ``REJECTED'' to ``PUBLISHED''. This shows that the GPM can be indicative of CVEs that are actively exploited, even if the CVE is marked as rejected.

\noindent \textbf{Other CVEs with the highest GPM scores.} As discussed previously, we collect and extract all CVE-IDs from GitHub exploit repositories in the dataset, regardless of their publication status. We now discuss some CVEs with high GPM scores in our dataset that did not have a published status in other sources:
\begin{itemize}[leftmargin=*, nosep]
    \item \textbf{CVE-2023-44976 ($GPM=117$):} The security researcher who discovered this vulnerability did not receive a response from the developers; however, on the day of the publication of the CVE, a ransomware group used this CVE in their attacks. After we provided the reports and GitHub repository to MITRE, they published this CVE on 2025-08-01.
    \item \textbf{CVE-2024-32640 ($GPM=79$):} The exploit was published on 2024-05-16. After we provided the exploit to MITRE and GitHub, they published the CVE on 2025-08-11. This CVE was added to VulnCheck KEV~\cite{vulncheck202432640cve} as it was detected by ShadowServer~\cite{shadowservers202432640cve} on 2024-05-16. 
    \item \textbf{CVE-2022-38694 ($GPM=496$):} This is the most popular CVE in our dataset with ``REJECTED'' status. We requested an update on this CVE from MITRE on 2025-08-26, and the CVE was published on 2025-09-01 (again, after the GPM picked up a signal about it).
    \item \textbf{CVE-2022-38691 ($GPM=81$):} Analyzed by the NCC group~\cite{Zhuravlev2022UnisocROMvuln}, the affected company signed up to be a CNA and rejected the CVE. On 2025-09-01, the CVE was published.
\end{itemize}

\noindent
Many vulnerability scanners and software composition analysis (SCA) tools use published CVEs. Similarly, the EPSS uses the same approach and only considers published CVEs. Our case studies show that the GPM is able to highlight important vulnerabilities even before they are noted by the common security software. As most tools and defenders are only aware of ``PUBLISHED'' CVEs, these tools would be at a disadvantage in such a case.    

However, the unpublished CVEs mentioned above are just examples of this risk. The 359 unpublished CVEs that we collected are only CVEs that were not published at the time of writing this paper. Based on the collected data, for \textbf{1,603} CVEs, the public GitHub exploit was available at least one day earlier than the date of publication of the CVE, and in \textbf{561} cases, the public exploit was published at the same time as the CVE was published. Thus, the attackers get an opportunity window having access to an exploit, where defenders are not even yet aware of the existence of such CVEs. Thus, adding the GPM to the set of indicators for prioritizing vulnerabilities can be very helpful for defenders in such scenarios.

Table~\ref{tab:case_studies} presents additional case studies on important CVEs that are flagged by the GPM. These case studies demonstrate the GPM's potential in uncovering critical exploited CVEs earlier than other data sources.

\begin{table*}[t]
\centering
\caption{Additional CVE case studies where the GitHub repository with an exploit proof-of-concept was published before the CVE.}
\label{tab:case_studies}
\begin{tabular}{p{3.2cm}p{11.2cm}}
\toprule
\textbf{CVE} & \textbf{Timeline / Notes} \\
\midrule
\addlinespace
CVE-2021-4034 & RedHat published on 2022-01-25; MITRE/NVD published on 2022-01-28. 
Between 2022-01-25 and 2022-01-27: 68 repositories with a GPM score of 2,074.  
Exploited by ransomware and botnets; detected in June 2022 and added to CISA KEV. \\
\addlinespace
CVE-2021-40444 & Microsoft disclosed on 2021-07-09; MITRE published on 2021-09-15.  
On 2021-09-08 to 2021-09-09, repositories with in-the-wild exploits appeared; the first weaponized repository appeared on 2021-09-10 with a GPM score of 1,397 \emph{before} MITRE publication.  
Exploited by ransomware; added to CISA KEV. \\
\addlinespace
CVE-2022-22965 (Spring4Shell) & Empty repository appeared on 2022-03-29; first PoC appeared on 2022-03-30, most likely from a 0-day from vx-underground samples; Spring.io advisory published on 2022-03-31; MITRE published on 2022-04-01; added to CISA KEV on 2022-04-04.
GPM score of 600 before CVE publication. \\
\addlinespace
CVE-2022-2588 & Found at Pwn2Own event. Linux fix submitted on 2022-08-09; ZDI publication on 2022-08-18; 
first PoC appeared on 2022-09-20; CVE published on 2024-01-08.
The long delay for the CVE publication is unclear. \\
\addlinespace
CVE-2022-40684 & Published by Fortinet on 2022-10-10; added to CISA KEV on 2022-10-11; MITRE published on 2022-10-18. The exploit was reverse-engineered from the patch\cite{Horseman2022FortiAuthBypass}. 398 GitHub stars. Exploitation through ransomware was confirmed~\cite{stairwell_akira_ransomware}. \\
\addlinespace
CVE-2025-0411 (7zip) & Public release in ZDI on 2025-01-19; PoC appeared on GitHub on 2025-01-22;
MITRE published on 2025-01-25; added to CISA KEV on 2025-02-04.
73 GitHub interactions before MITRE's publication. \\
\addlinespace
CVE-2024-25600 \linebreak(WordPress theme RCE) & CVE was reserved on 2024-02-08; added to VulnCheck KEV (PatchStack\footnote{\url{https://patchstack.com/articles/critical-rce-patched-in-bricks-builder-theme/}}) on 2024-02-19; exploit published on GitHub on 2024-02-20; CVE published on 2024-06-04.
215 GitHub interactions before publication of the CVE. \\
\addlinespace
CVE-2024-45519 \linebreak(Zimbra RCE) & CVE was reserved on 2024-09-01; a blog post was published on 2024-09-04; an exploit was published on 2024-09-28; added to VulnCheck KEV on 2024-10-01; CVE published on 2024-10-02; added to CISA KEV on 2024-10-03.
There were 42 interactions with the exploit repository before the publication of the CVE.
\\
\bottomrule
\end{tabular}
\end{table*}

\takeaway{4.2}{The GPM can identify exploited CVEs overlooked by existing sources, sometimes providing earlier signals of risk.}

\section{Discussion}\label{sec:discussion}

\textbf{Advantages and disadvantages of the GPM.}
The GPM has several important advantages. When compared to CISA KEV CVEs, the GPM has predictive power rather than being reactive, resulting in less lag for defenders. We have demonstrated its utility by, for example, identifying a falsely rejected CVE that was exploited by a ransomware group and highlighting inconsistencies in the SSVC labels. Checking exploit popularity on GitHub can be done systematically and automatically; thus, unlike CISA KEV, it will not be limited to manual verification by humans. CVEs published by vendors may take more time to be published by MITRE. The GPM is not affected by this delay, and unlike the EPSS, it is not limited to fully published CVEs. Most importantly, the GPM is a free, publicly available metric with high correlation to KEV CVEs. Therefore, it can be readily used by all practitioners and researchers interested in prioritizing CVEs based on exploitability and integrated into other available CVE scoring methods, such as the EPSS. 

On the other hand, because this metric is decentralized, there is no guarantee that all CVEs will be available on GitHub, so the GPM's CVE coverage is very limited: as we have seen, only 2.5\% of CVEs have PoC repositories. Moreover, the GPM can be affected by GitHub's indexing lag. Although our manual tests showed repository indexes on the same day, GitHub remains a 3rd-party service and is therefore prone to downtime and instability, even though this is unlikely given its high reputation.

We shared our dataset and a description of the GPM with the EPSS team to consider if it can be added as a feature source in the underlying machine learning model.‌ As a result, the GPM was integrated into the EPSS version 5. Due to technical reasons, the GPM integrated into the EPSS is modified and counts total stars, forks, and watches without considering the uniqueness of users.


\takeaway{5}{The GPM is based on publicly available data and is both accessible and reproducible for anyone. It can flag important CVEs early. Its key disadvantage is the relatively low coverage of CVEs.}

\noindent \textbf{Limitations.} 
We discuss the key limitations of our work.

\noindent \emph{KEVs.} To evaluate our results, we had a limited set of known exploited vulnerabilities. For example, in the EPSS paper~\cite{jacobs2023enhancing}, the researchers collected 12,243 exploited CVEs, whereas we collected only 4,610 due to the limited public availability of high-quality exploits. Thus, the threshold set for GPM can be changed. This is also true for AlienVault OTX. OTX contains much more data, but, as it was not possible to systematically use it as a KEV source, the included CVEs are limited; thus, in our experiments, this source suffers from False Negatives (i.e., CVEs that are actually exploited but missed in our data). 
It is important to note that many CVEs may have private exploits or be in sources that we did not consider. Moreover, vulnerabilities are sometimes exploited before receiving a CVE. Similar to other metrics such as KEVs and the EPSS, the proposed metric will lag in such cases, highlighting the CVE reservation and publication time as bottlenecks.

\noindent \emph{Data availability.} Some works, such as the EPSS, benefit from private exploit intelligence sources. The lack of public accessibility of such sources limits the quality of comparisons and experiments. This is also the case for the ``Expected Exploitability'' work, where the web application is no longer available, the code and data are not publicly available, and we were unable to contact the authors~\cite{suciu2022expected}.

\noindent \emph{Noisy data.} Another limitation of this work is the noisy and incomplete data. Since the GPM uses publicly available GitHub repositories, some of them may contain irrelevant information, malware, or fake exploits~\cite{elyadmani2025securepoc}. Our examination of a random sample of repositories (as discussed in Sec.~\ref{sec:data}) has shown that the noise level is relatively low. 
However, it is not possible to fully eliminate such repositories from our dataset. Increasing the threshold is expected to reduce the likelihood of such false-positive CVEs, since, by the GPM definition, a single user can increase the metric by at most one point. If a CVE exploit repository is popular but lacks exploit code, it can still indicate the CVE's popularity; thus, for this research, it remains a valuable indicator of the CVE's popularity. In the worst case, malicious actors may assign fake stars to many CVEs to inflate the popularity of unimportant CVEs and consume defenders' time. Stricter GitHub controls, including bot detection, real human identification, account age verification, and activity monitoring, may help reduce false positives. Moreover, some repositories may serve as a simple PoC rather than a working exploit, yet we did not distinguish them, as they still indicate community interest in a CVE. 
Other data sources we used are also prone to noise. For example, Dong et al. have shown inconsistencies between the CVE and NVD data~\cite{dong2019towards}, and, as our study shows, some CVEs can be misinterpreted by CNAs due to the complexity of the ecosystem.

Part of the research is based on the‌ history of stars received over time in GitHub repositories. As we finalized this work, GitHub introduced a change in their policy \footnote{https://github.com/orgs/community/discussions/202114}, preventing users from checking which user IDs starred a repository. Although the community is opposing the decision, it is unclear if the decision will be reverted.‌ In case the decision is kept as is, the alternative way to use GPM for newer data points is to use it without calculating unique users.‌ This change, however, may introduce different results in experiments, which is not considered in the presented analysis.

\section{Conclusions}\label{sec:conclusions}

Vulnerability prioritization is an important problem. In this work, we propose a new metric to prioritize likely exploited vulnerabilities. Our metric, called the GPM, is based on the popularity of CVE exploits published on GitHub, and it is open-source, transparent, and free for everyone. We have shown that it is independent of the known exploited vulnerability lists like CISA KEV and provides tangible insights into important CVEs that might be overlooked by other systems. The GPM insights have led MITRE to change the status of several CVEs to published, making them more visible to defenders. Moreover, the GPM was merged into the EPSS version 5 with modifications. Similar to KEV lists, while the CVE coverage of the GPM is limited, its ability to flag important CVEs and open and publicly available methodology makes it very useful to security practitioners.


\bibliographystyle{unsrt}
\bibliography{references}
\section*{Ethical Considerations}
We elaborate on our ethical considerations, which were informed by the Menlo report's principles~\cite{bailey2012menlo}.

\textbf{Beneficence}. 
The key stakeholders in our research are third-party system owners (whose systems may contain exploitable vulnerabilities), GitHub PoC repository owners, GitHub and other CVE-related data owners (e.g., CISA KEV, the NVD), adversaries seeking exploitation recipes, and our research team. We now analyze the potential harms to these stakeholders and weigh them against potential benefits.  
In this research, no third-party systems were penetration-tested, exploited, or examined in an intrusive manner. Thus, there was no direct risk to third-party system owners. To limit potential future harms to these stakeholders, we will limit the release of the PoC data, as specified below. Third-party system owners often prioritize vulnerabilities in their systems based on their exploitability. The GPM introduces a new, accessible, and repeatable vulnerability prioritization method for third-party system owners. Therefore, the insights of the proposed GPM metric offer substantial potential benefits to these stakeholders. 

We worked with open-source project data from GitHub. Mining software repositories entails some potential harms~\cite{gold2022ethics}. To mitigate risks to GitHub repository owners whose data are included in our dataset, we store the collected PoC repository data securely and will limit the release of the collected data. We thus consider the potential harms to this stakeholder group to be limited. Some of these repository owners might be security researchers or defenders interested in protecting against vulnerability exploitation~\cite{elyadmani2025securepoc}. These individuals will benefit from our study by learning about the GPM metric.

In our study, we used only data from publicly available, free sources. 
Exploits and PoCs, by nature, are dual-use content (i.e., can be used for adversarial activities), but we only examine properties of publicly available GitHub-hosted exploits (permitted under the GitHub terms of service~\cite{githubexploit}), without running them. We use the standard GitHub APIs to collect our PoC dataset. Similarly, we used dedicated public interfaces to collect data from our other sources, and we believe the resource consumption (as a potential harm) incurred by our data providers was very limited. 
To limit the potential benefits to adversaries seeking to exploit vulnerabilities, we will restrict the release of our dataset, even if the data is already publicly available. Furthermore, our GPM web service may enable attackers to observe highly exploited vulnerabilities. Since several other public data sources of exploited vulnerabilities exist (e.g., the KEVs or the EPSS) and, exactly like these sources, the GPM does not aid the miscreants in finding vulnerabilities in third-party systems, we consider the additional risk introduced by the GPM web service to be negligible.
We do not see significant risks to our research team. We were not exposed to harmful or very sensitive information or media. We benefit from this study by satisfying our research curiosity and our desire to help defenders prioritize relevant vulnerabilities. 

\textbf{Respect for persons.} While our study does not involve direct participants, we diligently protect the rights of the involved persons (GitHub repository owners): we collect only publicly accessible data from GitHub and, recognizing potential privacy concerns in repository data, we store it securely and limit its release.

\textbf{Respect for law and public interest.} We conducted our research respecting the relevant laws and regulations, specifically the EU GDPR. 

\textbf{Justice.}
We believe that the benefits and potential harms arising from our research are evenly distributed, and no stakeholder group bears disproportionate risk. We sought to minimize potential harms by following recommended best practices in security research (respecting GitHub's terms of service, protecting the privacy of GitHub repository owners, and limiting the release of our dataset).

\section*{Open Science}
\begin{figure}[t]
  \centering
  \includegraphics[width=1.00\linewidth, keepaspectratio]{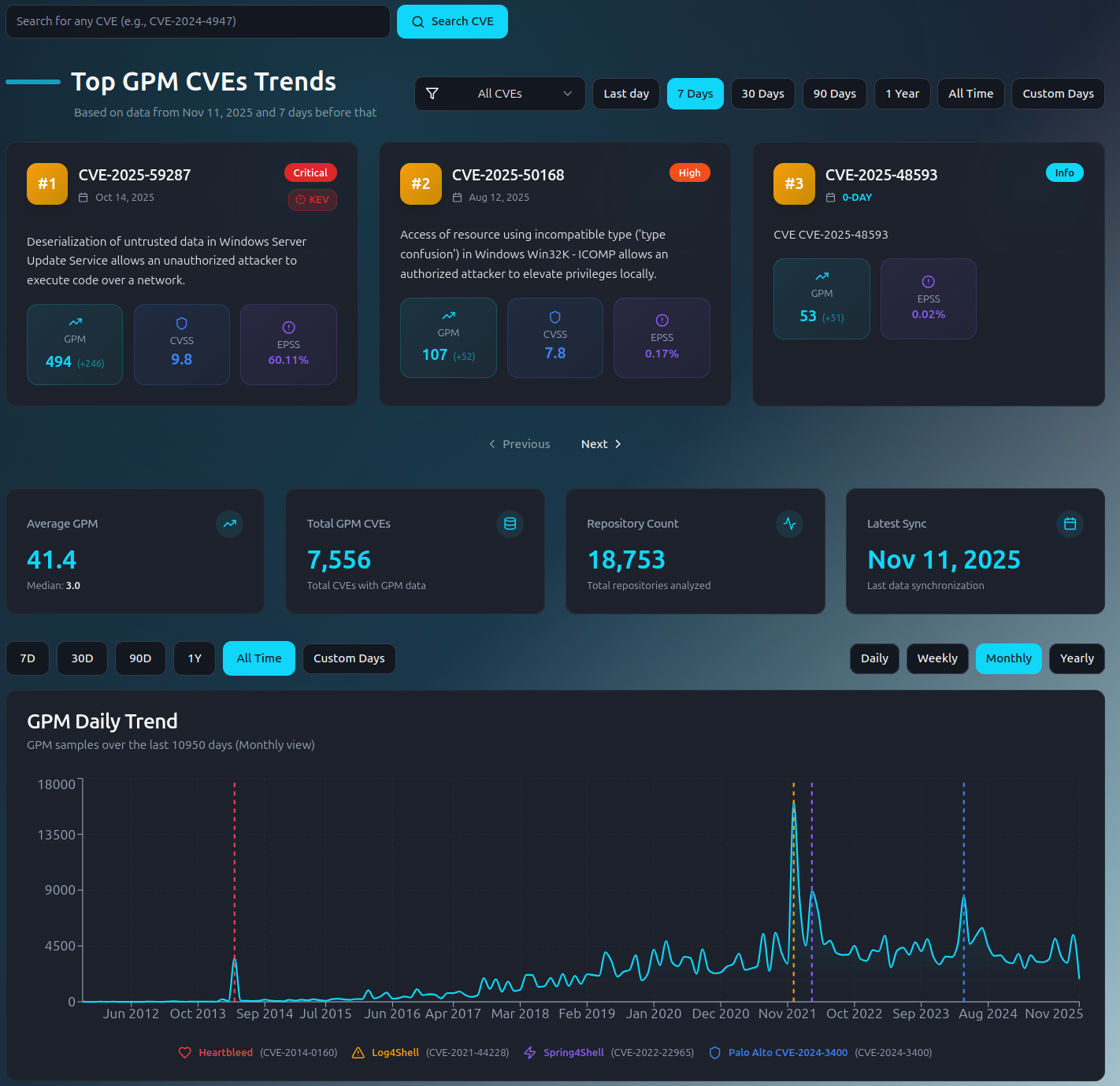}
  \caption{Web panel based on GPM data.}
   \label{fig:gpm-website}
\end{figure}

\arxivonly{
}
\reviewonly{
The source code that allows scraping all the discussed data is available at \href{https://anonymous.4open.science/r/gpm-f2EFmFhCa/}{https://anonymous.4open.science/r/gpm-f2EFmFhCa/}. We also provided a data dump and code to reproduce the experiments.
To increase the utility of GPM, we provide both a locally runnable version and an online website with the GPM scores for CVEs. A public but password-protected version is available at \url{http://45.14.224.176/}.
For review purposes, please use the following credentials: username:Moreover, we have attached a Docker image that allows running the website locally for review purposes with ease.
}

We intend to publicly share our source code that allows for collecting data in a reasonable time and calculating the GPM values. We will share our dataset responsibly with other researchers, based on individual verified requests. We will also provide a Docker image to run the GPM platform locally. A screenshot of the platform UI is shown in Figure~\ref{fig:gpm-website}. 

\section*{Disclaimer}
The data sources referenced in this work, including any names, trademarks, service marks, or logos, are used solely for descriptive and identification purposes. The respective organizations have not participated in, reviewed, or endorsed the preparation, findings, or conclusions of this work. All trademarks and logos are the property of their respective owners.

\appendix

\end{document}